\documentclass[]{aa}  

\usepackage{graphicx}
\usepackage{txfonts}
\usepackage{subfig}
\usepackage{textgreek}
\usepackage{multirow}
\usepackage{mathtools}
\usepackage{color}
\usepackage{lmodern}
\usepackage{epstopdf}

\bibpunct{(}{)}{;}{a}{}{,}

\newcommand{\FIG}[1]{}

\def\mso{\,{\rm M}_\odot}

\def\lso{\,{\rm L}_\odot}

\def\kms{\, {\rm km}\, {\rm s}^{-1}}

\begin{document}

% ***********************************************************************************************************************************************
%                                                                   TITLE PAGE
% ***********************************************************************************************************************************************

   \title{ALMA observations of the nearby AGB star L$_{\rm 2}$ Puppis}

   \subtitle{II. Gas disk properties derived from $^{\rm 12}$CO and $^{\rm13}$CO $J=$3$-$2 emission}

   \author{Ward Homan
          \inst{1}
          \and
          Anita Richards
          \inst{4}
          \and
          Leen Decin
          \inst{1}
          \and
          Pierre Kervella
          \inst{2,3}
          \and
          Alex de Koter
          \inst{1,5}
          \and
          Iain McDonald
          \inst{4}
          \and
          Keiichi Ohnaka
          \inst{6}
          }

   \offprints{W. Homan}          
          
   \institute{$^{\rm 1}\ $Institute of Astronomy, KU Leuven, Celestijnenlaan 200D B2401, 3001 Leuven, Belgium \\
	     $^{\rm 2}\ $Unidad Mixta Internacional Franco-Chilena de Astronom\'{i}a (CNRS UMI 3386), Departamento de Astronom\'{i}a, Universidad de Chile, Camino El Observatorio 1515, Las Condes, Santiago, Chile. \\
	     $^{\rm 3}\ $LESIA (UMR 8109), Observatoire de Paris, PSL Research University, CNRS, UPMC, Univ. Paris-Diderot, 5 Place Jules Janssen, 92195 Meudon, France \\
             $^{\rm 4}\ $JBCA, Department Physics and Astronomy, University of Manchester, Manchester M13 9PL, UK \\
             $^{\rm 5}\ $Sterrenkundig Instituut `Anton Pannekoek', Science Park 904, 1098 XH Amsterdam, The Netherlands\\
             $^{\rm 6}\ $Universidad Catolica del Norte, Instituto de Astronomia, Avenida Angamos 0610, Antofagasta, Chile \\
             }

   \date{Received <date> / Accepted <date>}
 
   \abstract  
   {The circumstellar environment of the AGB star L$_{\rm 2}$ Puppis was observed with ALMA in cycle 3, with a resolution of $15 \times 18 \rm\ mas$. The molecular emission shows a differentially rotating disk, inclined to a nearly edge-on position. In the first paper in this series (paper I) the molecular emission was analysed to accurately deduce the motion of the gas in the equatorial regions of the disk. In this work we model the optically thick $^{\rm 12}$CO $J=$3$-$2 and the optically thin $^{\rm 13}$CO $J=$3$-$2 rotational transition to constrain the physical conditions in the disk. To realise this effort we make use of the 3D NLTE radiative transfer code {\tt LIME}. The temperature structure and velocity structure show a high degree of complexity, both radially and vertically. The radial H$_{\rm 2}$ density profile in the disk plane is characterised by a power law with a slope of $-3.1$. We find a $^{\rm 12}$CO over $^{\rm 13}$CO abundance ratio of 10 inside the disk. Finally, estimations of the angular momentum in the disk surpass the expected available angular momentum of the star, strongly supporting the indirect detection of a compact binary companion reported in paper I. We estimate the mass of the companion to be around 1 Jupiter mass.}
%    {bla1}    
%    {bla2}    
%    {bla3}    
%    {bla4}
%    {bla5}
   

   \maketitle

% ***********************************************************************************************************************************************
%                                                                   END TITLE PAGE
% ***********************************************************************************************************************************************

% ===================================================================================================================================================

\tableofcontents

\section{Introduction}

Low and intermediate mass stars evolve up the asymptotic giant branch (AGB) as they reach the end of their lives. This phase is characterised by the development of a dense and cool molecular envelope scattered with (sub-)microscopic dust particles, which the star expels into space. Recent observations of the circumstellar environments (CSEs) of AGB stars with high-resolution instruments have revealed a rich spectrum of structural complexities. These include bipolar structures \citep[e.g.][]{Balick2013}, arcs \citep[e.g.][]{Decin2012,Cox2012}, shells \citep[e.g.][]{Mauron2000}, clumps \citep[e.g.][]{Bowers1990,Ohnaka2016,Khouri2016}, spirals \citep[e.g.][]{Mauron2006,Mayer2011,Maercker2012,Kim2013}, non-rotating tori \citep[e.g.][]{Skinner1998}, and bubbles \citep{Ramstedt2014}. The most significant source of asphericity in evolved stars are thought to be magnetic fields \citep[e.g.][]{Sanchez2013,VanMarle2014} and binarity \citep{Soker1997,Huggins2007}. The importance of the latter phenomenon cannot be understated, as the multiplicity frequency of the main sequence (MS) progenitors of AGB stars is found to exceed 50 percent \citep{Raghavan2010,Duchene2013}. 

\begin{figure*}[htp]
\centering
 \centering
 \resizebox{7.0cm}{!}{\includegraphics{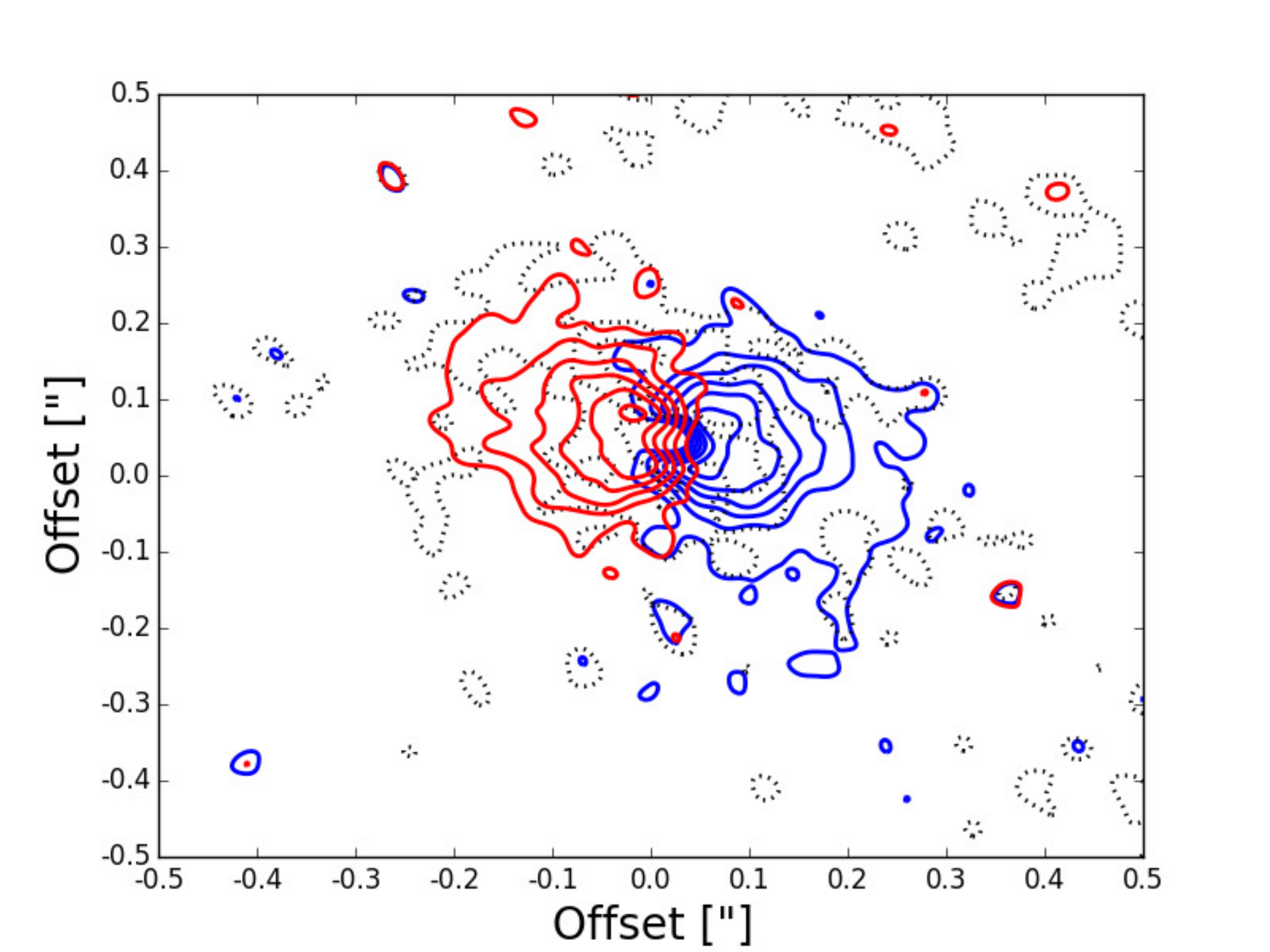}}
 \resizebox{7.0cm}{!}{\includegraphics{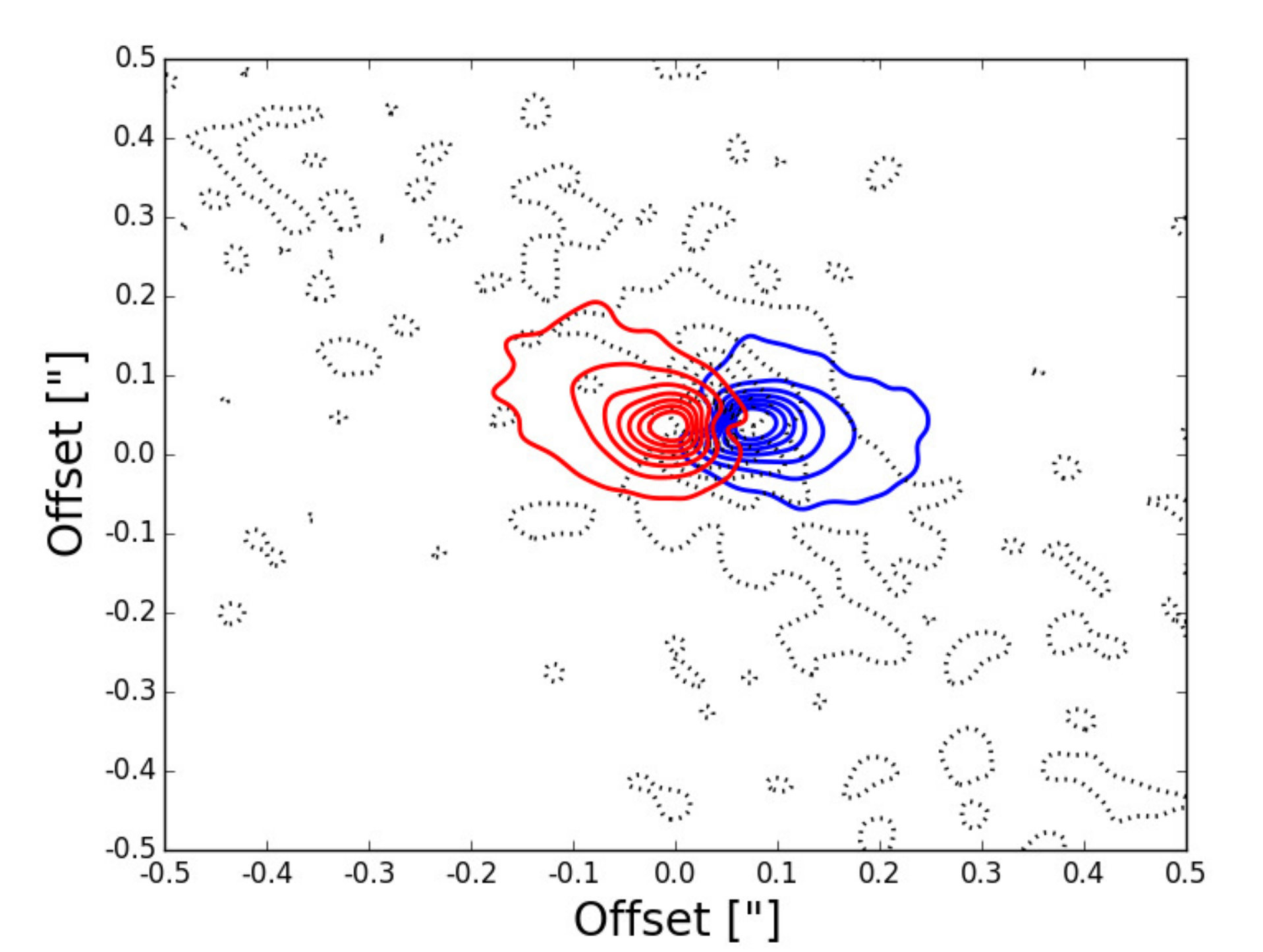}}
\caption{Stereogram view of the continuum subtracted $^{\rm 12}$CO $J=$3$-$2 (\emph{left panel}) and $^{\rm 13}$CO $J=$3$-$2 (\emph{right panel}) data. The blue (red) contours correspond to the velocity-averaged emission profile of 98\% of the blue-shifted (red-shifted) wing of line. The black dashed contours correspond to the velocity-averaged emission of the central 2\% of the emission line. The contours are drawn at every 2.5 times the continuum rms noise value ($^{\rm 12}$CO: 2.5 mJybeam$^{\rm -1}$, $^{\rm 13}$CO: 2.6 mJybeam$^{\rm -1}$). This figure appeared in \citet{Kervella2016}, and is repeated here as a guide. \label{dop_dat}}
\end{figure*}

L$_{\rm 2}$ Puppis is a semi-regular pulsating variable ($P=141$\,days, \citep{Kholopov1985, Bedding2005}) whose M5III spectral type corresponds to an effective temperature of $T_\mathrm{eff} \approx 3500$\,K. It has radial velocity relative to the Local Standard of Rest (lsr) of $v_{\rm lsr} = 33.3 \kms$. At a distance of only 64\,pc ($\pi = 15.61 \pm 0.99$\,mas; \citep{VanLeeuwen2007}) L$_{\rm 2}$ Pup is the second nearest asymptotic giant branch (AGB) star next to R\,Doradus ($\pi = 18.31 \pm 0.99$\,mas), and it is $\approx 30\%$ closer than \object{Mira}. Recent continuum observations by \citet{Kervella2014,Kervella2015} show a compact dust disk around the central star, with an inner rim of 6 AU and an outer edge of 13 AU, at an inclination of approximately 82$^\circ$.

In November 2015, we obtained spatially resolved ALMA band 7 observations of the CSE of L$_{\rm 2}$ Pup using the most extended, 16-km configuration (see Sect. 2). The data shows a differentially rotating gas disk, oriented in a nearly edge-on position, adding yet another morphology to the above list. The very long baselines allowed us to reach an angular resolution of $15 \times 18 \rm mas$. The high spectral resolution in the molecular data enabled us to accurately map the motion of the gas in the equatorial regions of the disk in the first paper in this series \citep{Kervella2016}. The Keplerian motion of the gas in the inner regions of the disk have been used to accurately determine the central mass of the L$_{\rm 2}$ Pup system. In addition, a significant asymmetry in the continuum emission of the east and west sides of the disk points towards the presence of a binary companion.

In the present work we model the observed emission of the $^{\rm 12}$CO and $^{\rm 13}$CO rotational transition $J=$3$-$2 around L$_{\rm 2}$ Pup, which allows us to make predictions on the (thermo)dynamical and morphological properties of the circumstellar gas. This provides valuable constraints on the possible physical or thermo-chemical formation mechanisms producing such a morphological complexity, as well as insights on the local physics that dominate the inner wind. In addition, revealing and understanding the local inner wind dynamics sheds light on the processes that dictate the further evolution into post-AGB stars and planetary nebulae, whose morphologies have been extensively documented and have been found to be highly aspherical. Therefore, an in-depth understanding of the finer physical details governing the inner regions of complexity-harbouring AGB CSEs will aid in the quest to understand the missing morphological link between AGB stars and their progeny.

\section{ALMA data description}

For an extensive description of the ALMA observing set-up as well as an in-depth discussion of the continuum residuals see \citet{Kervella2016}. We summarise the most relevant information here.

L$_{\rm 2}$ Pup was observed on 2015 Nov 5 by ALMA for project code 2015.1.00141.S (PI Kervella), starting at UT 08:39:25 (epoch 2015.8448, MJD = 57 331.361) and ICRS position RA 07:13:32.475914 DEC --44:38:17.91396. There were 45 antennas present for all or most of the observations, providing baselines from 0.09 to 16 km. However, the coverage inside 350 m is very sparse, giving a maximum angular scale of about 250 mas for reliable imaging. Standard ALMA observing and data reduction procedures were used, including phase-referencing. The flux scale is accurate to about 7\% and the absolute astrometric error is about 7 mas; comparisons within this data set are much more accurate, limited only by the signal-to-noise ratio. The line-free continuum channels of L$_{\rm 2}$ Pup were used to self-calibrate the continuum and the solutions applied to all target data, and then the continuum was subtracted. Image cubes were made for the lines present in the data, and were adjusted to constant systemic velocity $v_{\rm sys}$ in the target frame.

In this paper, we present the $^{\rm 12}$CO and $^{\rm 13}$CO cubes. These have a spectral resolution of $\sim 0.22 \kms$ and a rms noise (off-source) of 2.5 and 2.6 mJy beam$^{\rm -1}$, respectively. The synthesised beam size is $\sim (0\farcs020 \times 0\farcs015)$, where the exact value depends on frequency. A 2.5 mas pixel size and an image size of $2\farcs56$ was used.

In the following sections we describe the morphological properties of the gas emission, as deduced from velocity channel maps, and position-velocity diagrams (PVDs). We identify the patterns within the inner arcsec as consistent with the emission produced by a rotating disk of gas, and describe the fine features present in the data in detail. All descriptions relating to velocity are made relative to $v_{\rm lsr} = 33.3 \kms$, i.e. the estimated systemic velocity.

\begin{figure*}[htp]
\centering
 \centering
 \resizebox{8.5cm}{!}{\includegraphics{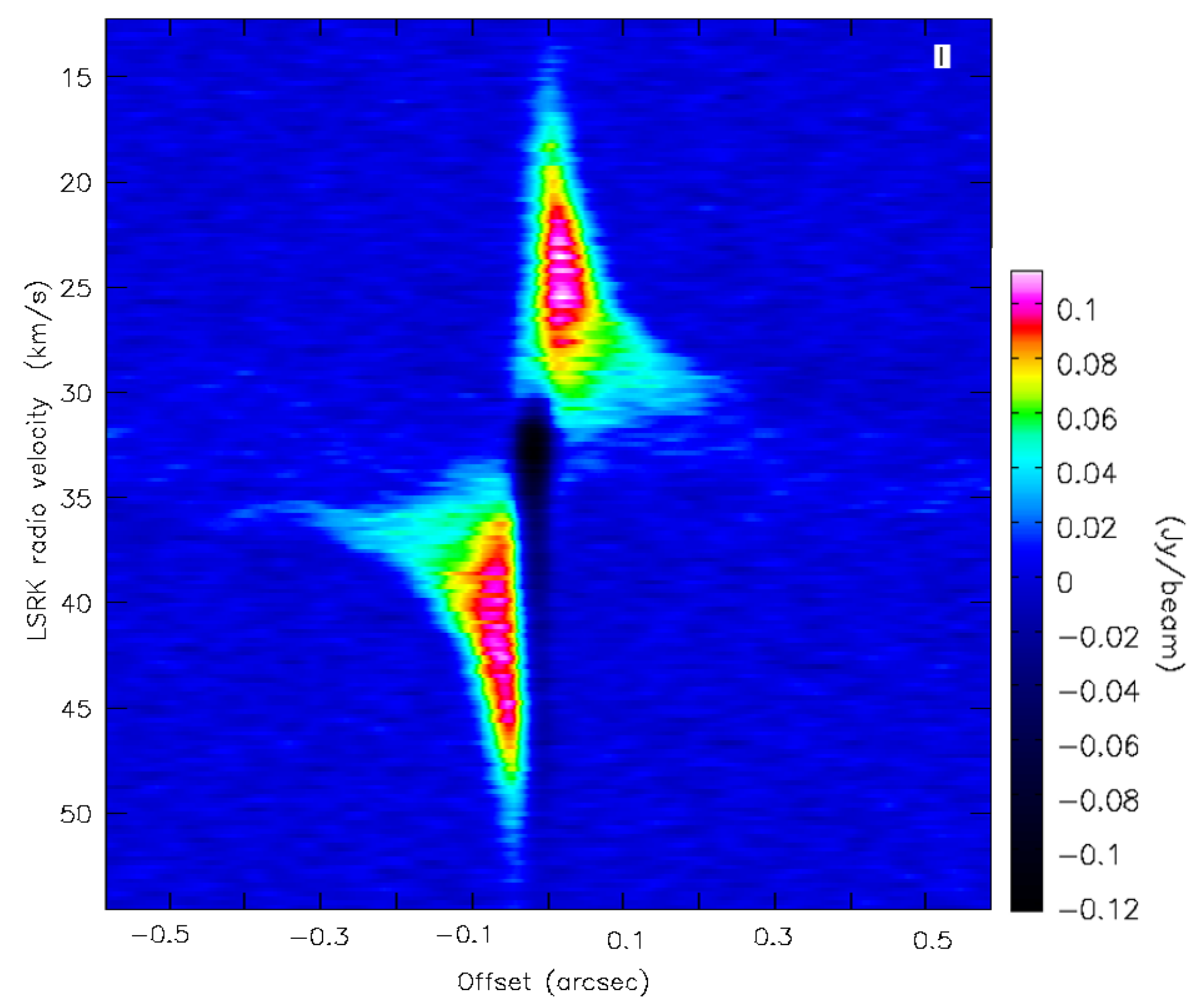}}
 \resizebox{8.5cm}{!}{\includegraphics{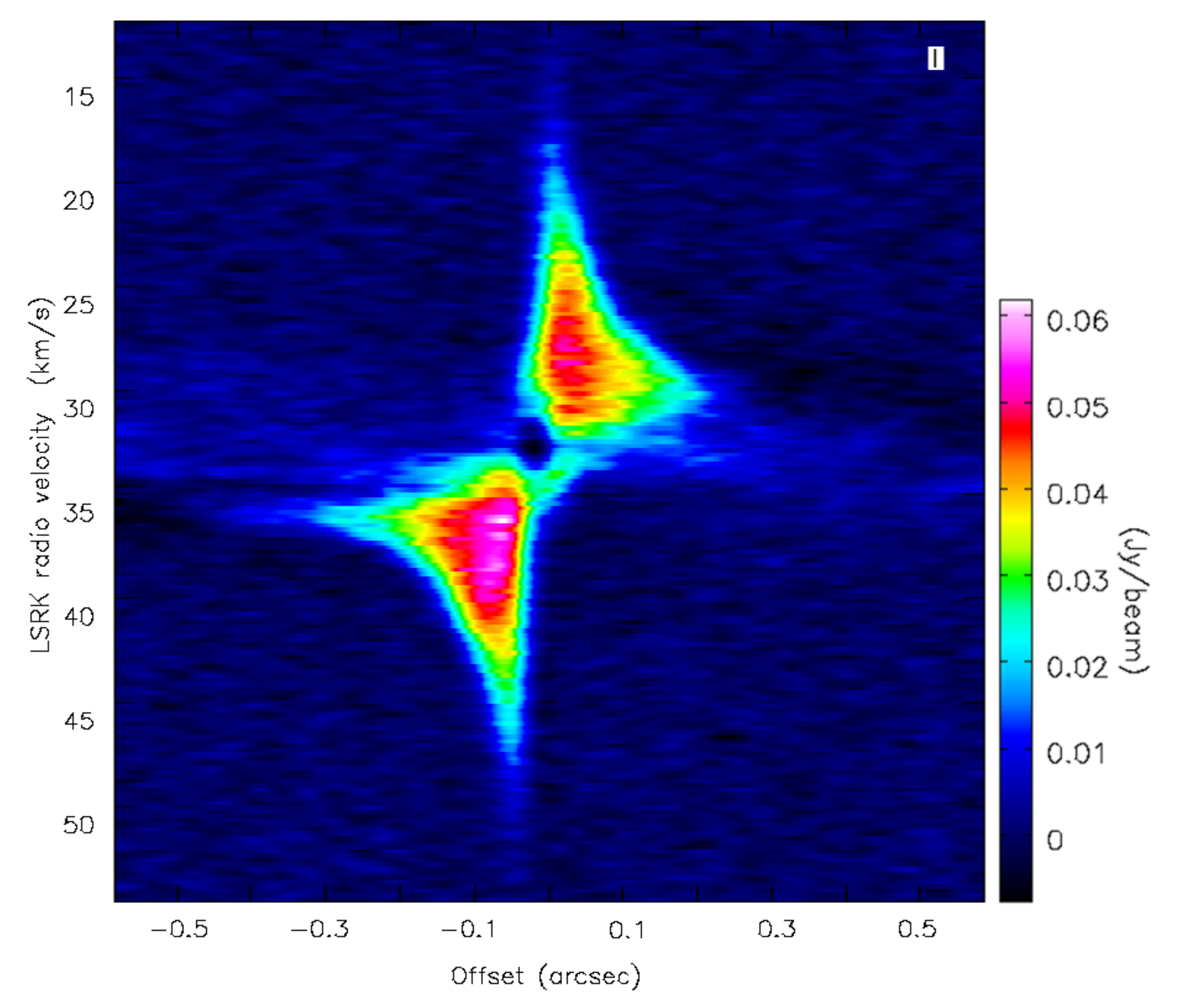}}
 \caption{Images of the PVD of the continuum subtracted $^{\rm 12}$CO data created with a slit parallel to the disk. \emph{Left:} PVD created with a minimal slit width of 15 mas. \emph{Right:} slit width equal to the height of the disk emission, which is 280 mas. For reference, $v_{\rm sys} = 33.3 \kms$.  \label{PV2_data}}
\end{figure*}

The intensity maps of the $^{\rm 12}$CO and $^{\rm 13}$CO emission as a function of velocity are visually represented in Figs. \ref{chan_contsub} and \ref{chan_contsub2}, respectively. The figures are comprised of 49 panels, selected from the data cubes at even intervals in velocity space. The velocity range of the panels span a domain selected such as to best show the spatial extent of the emission.

\subsection{$^{\rm 12}$CO $J=$3$-$2 emission}

The channel maps (Fig \ref{chan_contsub}) reveal horizontally elongated emission features, which range up to radial angular distances of 300 mas away from the central star, and vertical heights up to 300-350 mas. These features show a very clear east-west spatial offset separation between blue and red-shifted emission. In addition, the highest (absolute) velocities probe the regions closest to the central star, while the lower (absolute) velocities probe the outer regions. These features clearly point to a compact disk with a global velocity field that has the characteristics of differential rotation, seen edge-on. This claim is further supported by making integrated-intensity maps, or stereograms, of the data (Fig. \ref{dop_dat}), where the spatial and spectral separation between both sides of the disk are clearly apparent.

The high-velocity emission of the inner disk probes only a very small vertical region. Shifting to velocities closer to the stellar velocity the extent of this vertical signal increases. This suggests that the disk is flared. However, the emission does not seem to support a stereotypical flaring pattern (i.e. an increasing rate at which the scale-height grows with radius). Rather, the ratio of the scale height over distance seems to decrease with respect to distance from the centre.

Although expected to be present, the inner rim of the gas disk is not clearly visible in the data. This is mainly due to a combination of the ALMA beam size and the fact that the tilt of the disk obscures a significant portion of the inner rim. Nevertheless, the channel maps show a rather wide initial scale height, indicating that the inner rim of the gas disk may be relatively extended (of the order of a few AU) or even inflated or puffed up \citep{Dullemond2001,Dullemond2004}. 

\begin{figure}[htp]
\centering
\includegraphics[width=0.49\textwidth]{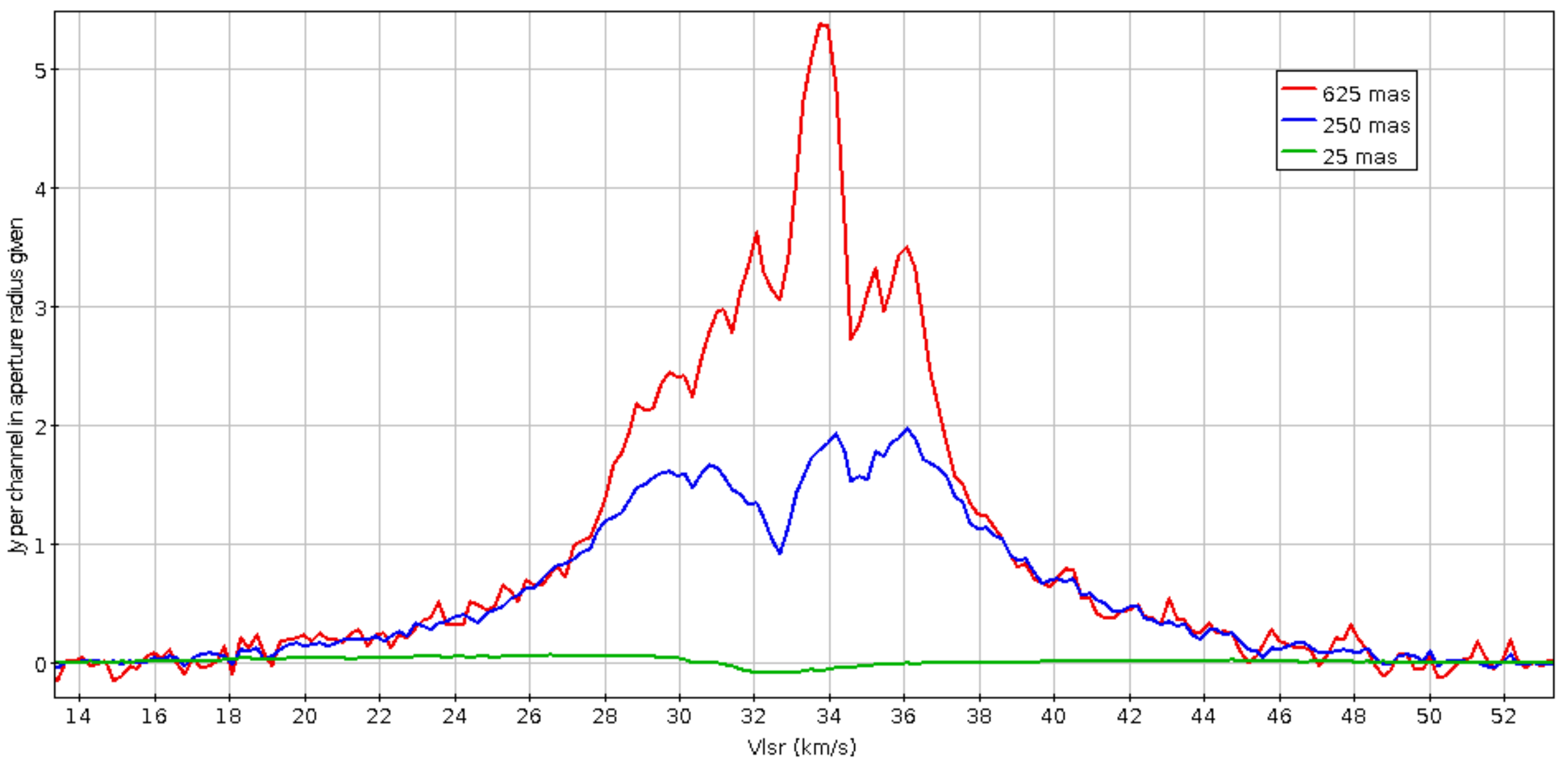}
\caption{Spectral lines of the $^{\rm 12}$CO J$=$3$-$2 emission, labelled by the different apertures used to extract them.  \label{line}}
\end{figure}

The emission around central velocity appears more diffuse and extended. This emission is probably dominated by the extended CSE of L$_{\rm 2}$ Pup. However, owing to the poor uv coverage of the smallest baselines of the configuration, length scales above 250 mas have been poorly sampled. In addition, the maximum recoverable scale of the configuration is $\sim$450 mas. Thus, some of the extended emission is probably resolved out and the remaining flux produces low-level artifacts, limiting the dynamic range in these channels. \citet{Kerschbaum1999} detected $^{\rm 12}$CO J$=$3$-$2 with the JCMT (beam size 13"), with a peak of $\approx59$ Jy (using S(Jy)/T$_{\rm mb}$(K) = 15.6), and estimated an expansion velocity of 3 $\kms$; their sensitivity was of order 1 Jy and the extreme-velocity wings and disk emission was not detected. Fig. \ref{line} shows spectra extracted from the ALMA data in apertures of the indicated radii. Thus, the JCMT dectected almost ten times as much emission as ALMA from the extended CSE.

Fig. \ref{line} also shows how, using a 250-mas aperture similar to the size of the disk (i.e. all the flux is recovered), there is a drop in intensity that is close to the systemic velocity, and using an aperture comparable to the size of the star itself shows absorption. This is seen in the corresponding channels of Fig. \ref{chan_contsub} where the emission becomes very faint. This is probably due to self-absorption along the line of sight to the star by the outer regions of the disk, which must therefore be optically thick and too cool to emit significantly at 345 GHz.

On smaller scales, which are visible on both the blue- (-4.5 to -3.0 $\kms$) and red-shifted (3.0 to 5.0 $\kms$) sides, the emission is slightly brighter above and below the midplane, compared to a narrow horizontal strip passing through the star. These features may be caused by local heating of the inner disk walls directly facing the stellar radiation field. This would also explain the puffed-up inner rim. We discuss this hypothesis in Sect. 6.2.1.

\begin{figure}[htp]
\centering
\includegraphics[width=0.45\textwidth]{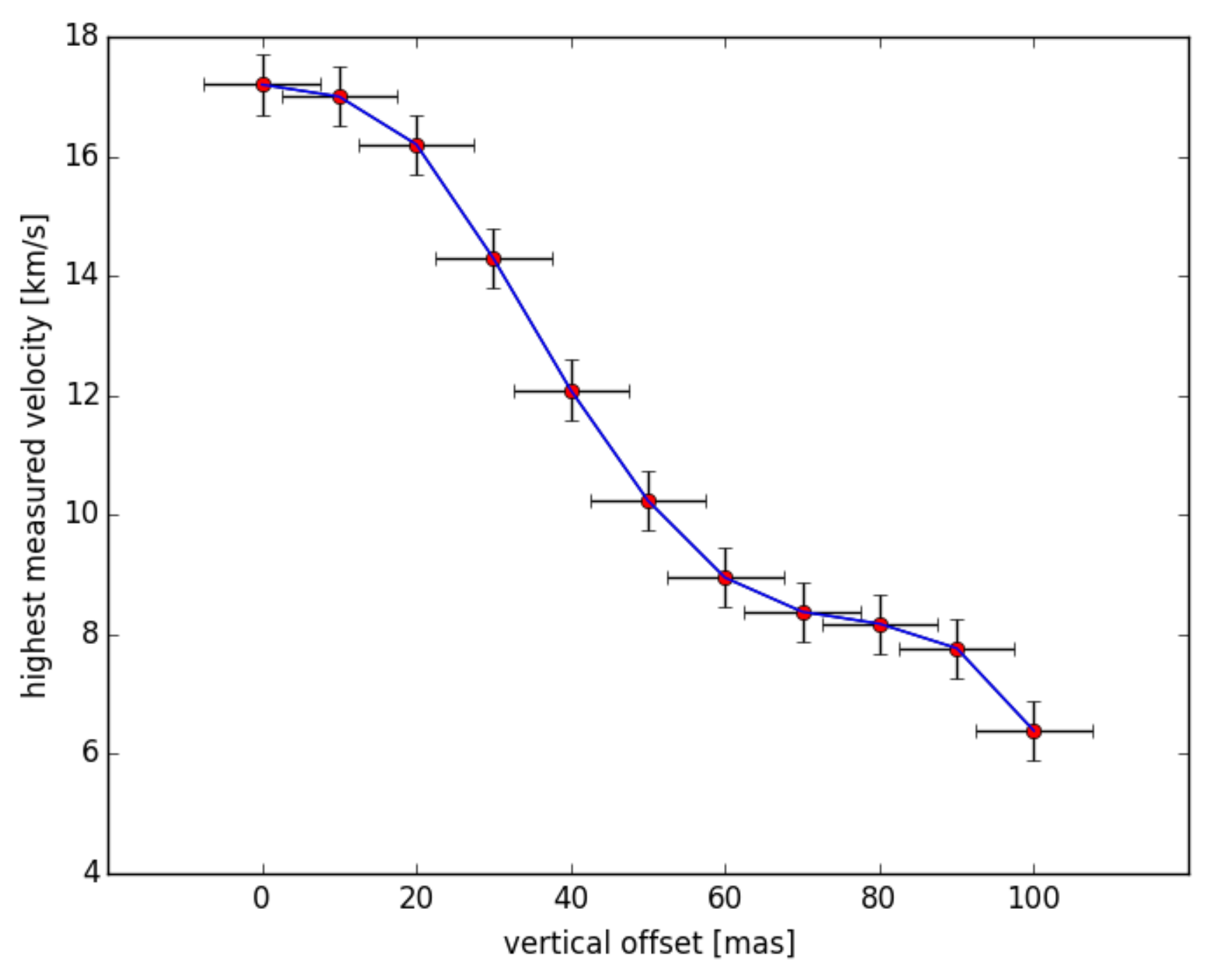}
\caption{Dependence of maximal gas velocity in the disk as function of vertical spatial offset deduced from the $^{\rm 12}$CO data. The uncertainty in velocity is 0.5$\kms$ (see Sect. 4.1.3); the spatial uncertainty is equal to the beam width.  \label{velOff}}
\end{figure}

The PVDs created by positioning the slit parallel to the disk plane (Fig. \ref{PV2_data}) permit the quantification of the spatial offset increase as a function of velocity. These trends have been extensively analysed by \citet{Kervella2016}, resulting in a very accurate determination of the central mass of the system (0.659 $\pm$ 0.052 $\mso$). The PVDs nicely exhibit the lack of compact emission around zero velocity. However, some diffuse emission can be seen around zero velocity up to large spatial offsets. This is the poorly sampled, large-scale emission discussed above. Comparing both diagrams it can be seen that the high-velocity tails do not span the same width in velocity space. The tail (of significant signal-to-noise) of the narrow slit PVD is longer by about 5$\kms$. This suggests that the material of the inner disk has higher tangential velocities close to its equator, compared to the gas at higher vertical offsets. To quantify this velocity gradient, we constructed 11 PVDs with a vertical separation of 10 mas. We extracted the maximal velocity at each offset by measuring the length of the tails of each PVD, employing the criterion that the signal must be twice the RMS noise value, outside the line. Indeed, this operation permitted us to recover the trend shown in Fig. \ref{velOff}. One possible cause for this vertical profile is that one may expect the upper disk regions to strongly interact with the AGB outflow, causing hydrodynamical perturbations that would impact the velocity. We discuss this further in Sect. 6.1.2.

\subsection{$^{\rm 13}$CO $J=$3$-$2 emission}

\begin{figure*}[htp]
\centering
 \centering
 \resizebox{8.5cm}{!}{\includegraphics{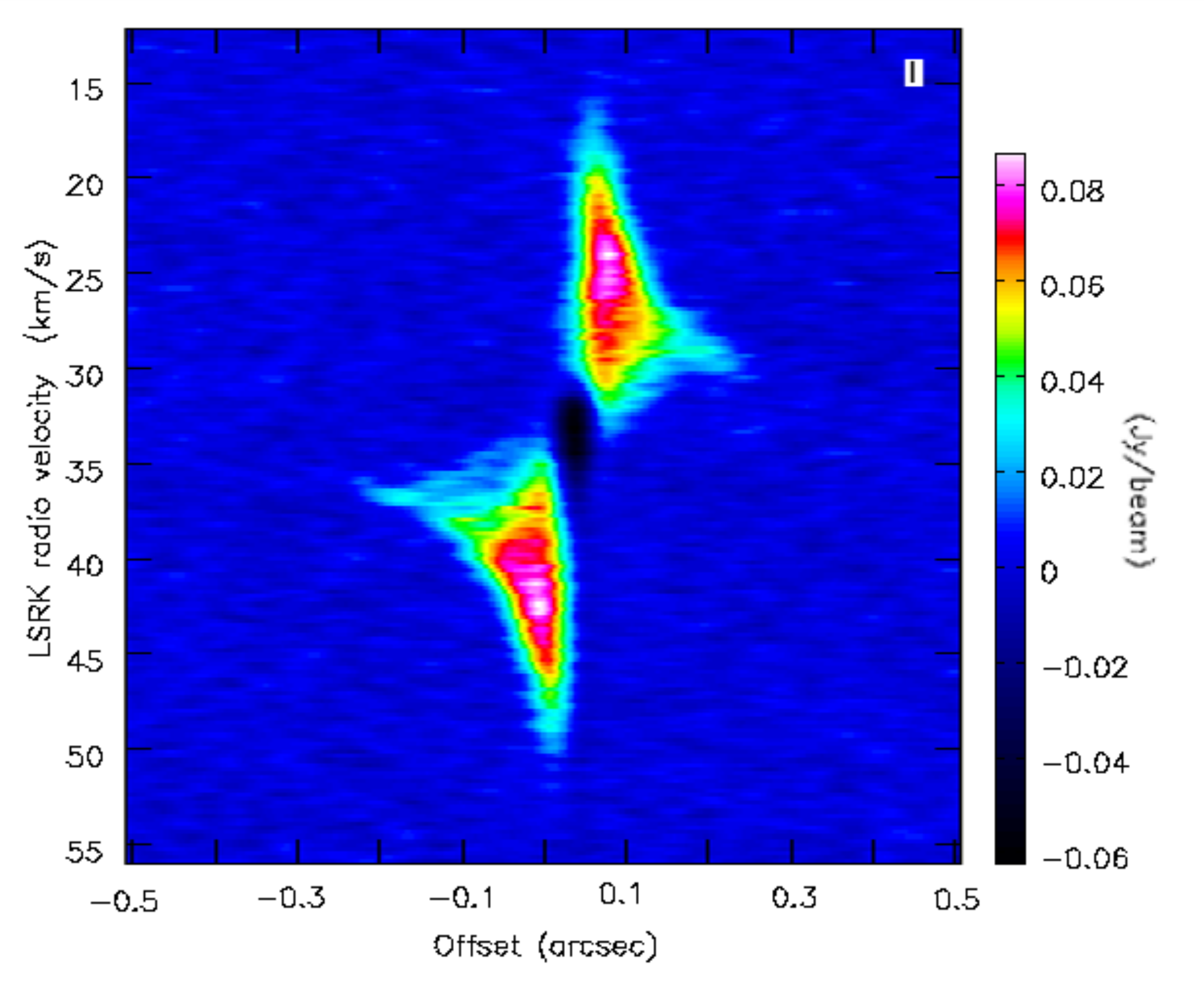}}
 \resizebox{8.2cm}{!}{\includegraphics{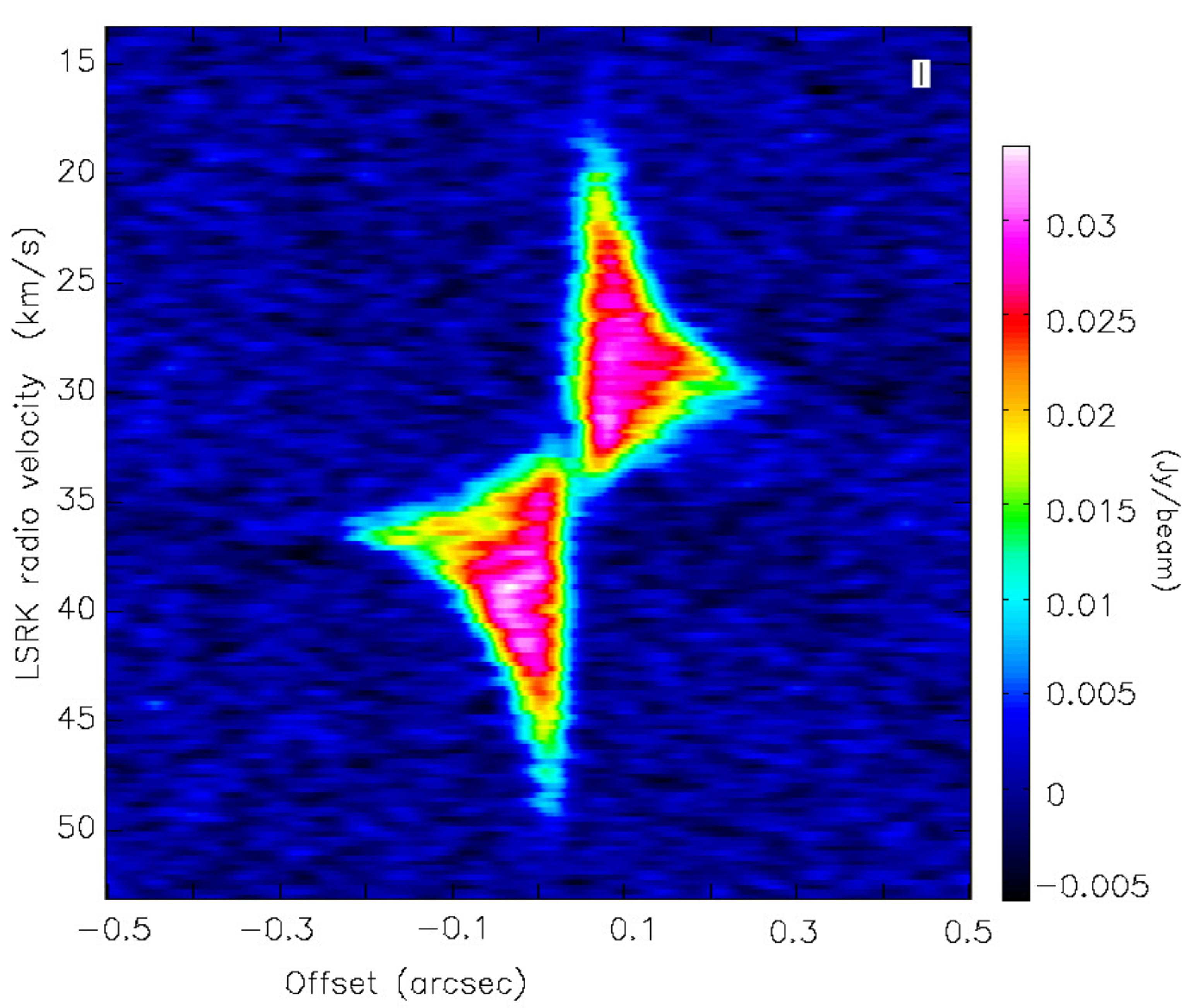}}
 \caption{Images of the PVD of the continuum subtracted $^{\rm 13}$CO data, created with a slit parallel to the disk. \emph{Left:} minimal slit width of 15 mas.  \emph{Right:} slit width equal to the height of the disk emission, which is 190 mas. For reference, $v_{\rm sys} = 33.3 \kms$.  \label{PV2_data2}}
\end{figure*}

Morphologically, the qualitative descriptions outlining the $^{\rm 13}$CO emission patterns (Fig \ref{chan_contsub2}) are in line with the $^{\rm 12}$CO emission (Fig \ref{chan_contsub}). In principle no major intrinsic differences are to be expected. The $^{\rm 13}$C/$^{\rm 12}$C ratio (see e.g. \citet{Hinkle2016} for estimates in AGB stars) is likely to be reflected in the $^{\rm 12}$CO/$^{\rm 13}$CO ratio, although not necessarily in a one-to-one relation because of isotope-selective photodissociation and chemical fractionation; therefore we expect the $^{\rm 13}$CO emission to have a lower optical depth than that of the $^{\rm 12}$CO emission. Hence, the absolute amount of observed emission is substantially lower in the $^{\rm 13}$CO emission, pushing the emission from the low-density regions of the disk below the sensitivity levels of ALMA. This is best seen in the apparent height of the $^{\rm 13}$CO disk, which is noticeably smaller than the $^{\rm 12}$CO disk (see Fig. \ref{dop_dat}). In addition, the $^{\rm 13}$CO disk is mostly optically thin, as it is still visible around the central velocity, contrary to the $^{\rm 12}$CO disk (see Fig. \ref{chan_contsub2}). Combined, these effects strongly dilute and blur the finer details present in the $^{\rm 12}$CO emission maps, resulting in a much smoother emission distribution.

There is, however, a strong asymmetry between the low-velocity blue-shifted side compared to the low-velocity red-shifted side. This asymmetry is best seen by comparing the emission in the -3.5 to -2.0 $\kms$ channels with the 1.0 to 3.5 $\kms$ channels. In these channels, the blue-shifted emission conforms with the expected patterns of a rotating disk. However, the red-shifted emission seems to be substantially perturbed, causing a deviation from the expected emission morphology. Emission around the equatorial regions does not seem to be present, while an emission tendril is seen to reach out to the north-east. This feature subsides beyond 3.5 $\kms$. Although one would expect the presence of a binary companion (\citep{Kervella2016}) to generate local perturbations, the projected velocities at which the perturbation is seen are too low to be related to the companion system, which, due to its proximity to the central star, can be expected to only substantially affect the highest velocities. We therefore interpret this feature as a hydrodynamical inhomogeneity in the outer boundaries of the disk.

The $^{\rm 13}$CO PVDs (Fig. \ref{PV2_data2}) show a high degree of similarity with the $^{\rm 12}$CO diagrams. In fact, the only clear difference is the length of the high-velocity tail, which seems to be identical for both the narrow-slit and the wide-slit PVDs. This can again be attributed to the molecular abundance fraction. The confined vertical extent of the $^{\rm 13}$CO emission means it mainly probes the Keplerian equatorial regions. The high-velocity tails of the PVDs of the $^{\rm 13}$CO emission are thus mostly insensitive to the sub-Keplerian velocities at higher vertical spatial offsets.

\section{Computational methods}

In this section we briefly describe the computational methods and strategy used to model the L$_{\rm 2}$ Pup CO line emission. We perform 3D radiative transfer via the {\tt LIME} code. The result of the radiative transfer calculations are subsequently subjected to the simulation observation algorithms of {\tt CASA} to generate 3D synthetic data that can be directly compared to the observed L$_{\rm 2}$ Pup data.

\subsection{LIME:}

We used the full 3D, non-local thermodynamical equilibrium (NLTE), infrared (IR) and submillimeter radiative transfer (RT) code {\tt LIME} for the radiation transport calculations; an in-depth description of the methods and algorithms utilised by the code can be found in \citet{Brinch2010}. The model is sampled by $10^5$ grid points, of which half are distributed logarithmically (gradually refining the grid towards its centre), and half are distributed randomly. In addition, the position of the grid points is weighted by relative density, further increasing the mesh refinement in the high-density regions of the model. Another $10^4$ grid points are positioned at the edge of the numerical domain, representing the points where the RT calculations are finalised. This total of $11\times10^5$ grid points are Delaunay-triangulated to form approximately $7 \times 10^5$ tetrahedral cells, which are subsequently Voronoi-tesselated and inside which the physical conditions are assumed to be constant. The physical set-up is described in terms of density, temperature, molecular abundance, macro- and micro-scale velocity fields. In order to detemine the mean intensity field, the RT equations are solved along the Delaunay lines connecting the neighbouring grid points up to the edge of the numerical domain. Once the RT calculations are finalised and the level populations have converged, the simulation is ray traced from a specific vantage point. It is important to note that the code solves the RT equations exclusively in order to converge the level populations for a given initial physical set-up. This user-defined physical set-up is thus not modified as the level populations converge. This, in turn, means that self-consistency of the physical model is only guaranteed if it was verified prior to the application of {\tt LIME}. A final remark is that {\tt LIME} was developed primarily for the modelling of molecular emission. The manner in which dust is taken into account in the calculations is rather rudimentary, and only suitable for first order dust modelling. However, because the CO molecule has such a low electric dipole moment, the contributions of the diffuse radiation field (from e.g. the dust) to the level excitation are not very important, diminishing the available free parameters and thus the modelling difficulty. The spectroscopic CO data of the LAMDA database \citep{Schoier2005} were used; the collisional rates were taken from \citet{Yang2010}. Because of its low electric dipole moment, the surrounding radiation field is not expected to affect the CO emission much. Nevertheless, both dust and the stellar radiation field were taken into account in the radiative-transfer calculations. The dust composition has been adopted from \citet{Kervella2015}, with the dust density distribution following the gas density distribution and a gas-to-dust mass ratio of 100. The stellar radiation field is approximated by a black body with a temperature of 3500K and a luminosity of 2500 $\lso$.

\subsection{CASA:}

After retrieving the intrinsic emission distribution from {\tt LIME}, we post-processed it with {\tt CASA} \citep{McMullin2007} to simulate an observation with ALMA. The actual general observation conditions and instrumental set-up were adopted as input parameters for the simulations, and are shown in Table \ref{casa}. This also implies we used the same antenna locations to give consistent visibility plane coverage between the observations and simulations.

\begin{table}[htp]
\centering
\begin{tabular}{ l l }

\hline
\hline
\multicolumn{2}{ c }{Simulation Parameters} \\
\hline
Pixel size of input model & 0.0047'' \\
Field size of input model & 1.25'' \\
Peak flux & Taken from {\tt LIME} output  \\ 
Transition 1 & $^{\rm 12}$CO $J=$3$-$2 (345.765 GHz) \\
Transition 2 & $^{\rm 13}$CO $J=$3$-$2 (330.587 GHz) \\
\hline
Pointing & Single \\
Channel width $^{\rm 12}$CO & 0.12MHz \\
Channel width $^{\rm 13}$CO & 0.22MHz \\
\hline
PWV & 0.42 mm \\
Thermal noise & standard \\
Ground temperature & 269 K \\
Integration time & 7953 sec on-source \\
\hline

\end{tabular}
\caption{ALMA observation simulation specifications, see the ALMA technical handbook. \label{casa}}
\end{table}

\subsection{Modelling strategy}

We wish to make it clear that our best-fit models are not achieved via the exploration and statistical analysis of a grid of models covering the more than 20 free parameters available to us. Such an analysis cannot simply be performed with present-day computational facilities. The best-fit model we obtained was created by making reasonable assumptions and educated guesses on the general physical properties (based on previous studies and the nature of the observed emission patterns), and subsequently attempting to converge to the reproduction of the emission characteristics via iterative adjustment of these parameters. Thus, when we use the term `best fit' in this paper, we do not refer to a statistically significant result that follows from extensive and detailed post-processing, but rather to the set of parameters for which the simulated emission shows a good visual resemblance to the emission patterns obtained by ALMA.

Owing to these computational limitations, we devised a strategy for the determination of the physical disk properties via RT modelling. We decided to start by focusing the analysis on the $^{\rm 12}$CO emission. The reasons for this which all stem from the expected lower $^{\rm 13}$CO molecular abundance compared to the $^{\rm 12}$CO abundance \citep{Hinkle2016}, are listed below:
\begin{itemize}
 \item $^{\rm 13}$CO over H$_{\rm 2}$ abundance ratios are not well known, while $^{\rm 12}$CO over H$_{\rm 2}$ ratios are better constrained.
 \item The lower abundance of $^{\rm 13}$CO results in a flatter and dimmer disk, in which the high degree of complexity in the thermodynamical, kinematical and morphological properties of the disk are less pronounced.
 \item Photodissociation effects, if present, have a larger impact on the overall morphology of the $^{\rm 13}$CO disk. We make no attempt to quantify the magnitude of photodissociation effects on the gas, so using the $^{\rm 12}$CO emission as a starting point for the modelling drastically reduces this systematic error.
 \item We can expect the $^{\rm 13}$CO emission to be optically thin, causing temperature and absolute density to be fully degenerate. Using the optically thick $^{\rm 12}$CO emission untangles both the temperature and the denisty by strongly reducing the influence of the absolute density.
\end{itemize} 
After we determined the set of physical conditions that best reproduce the $^{\rm 12}$CO emission patterns, we refined the model using the $^{\rm 13}$CO emission, making sure to retain consistency with the $^{\rm 12}$CO emission.

The extended emission around central velocity (visible in Fig. \ref{chan_contsub}, discussed in Sect. 2.1) is probably not part of the disk system, and we therefore do not include it in the modelling effort.

\section{Physical structure of the gas disk}

\subsection{Dominant and turbulent velocity field}

\begin{figure}[htp!]
\centering
\includegraphics[width=0.45\textwidth]{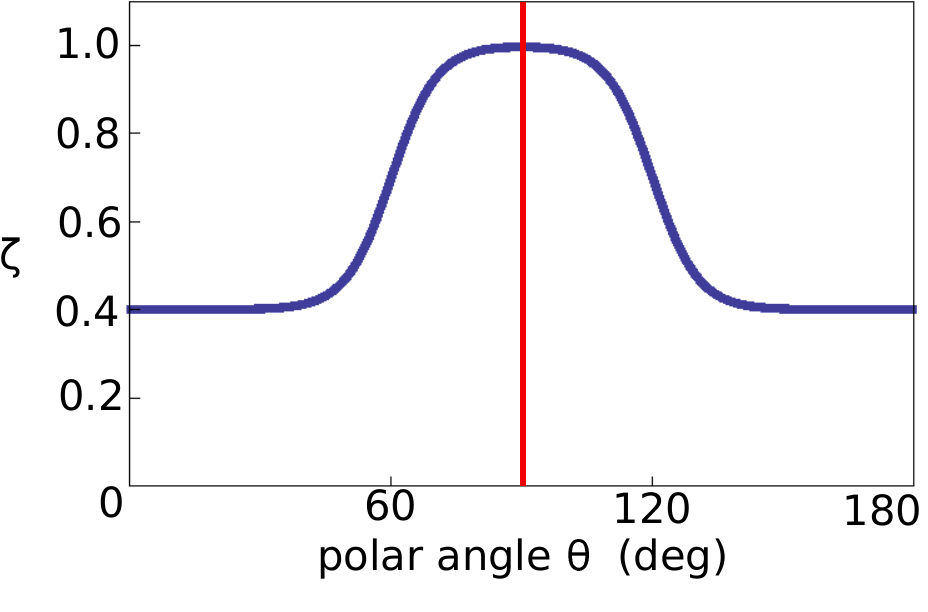}
\caption{Rate, $\zeta$, at which the equatorial tangential velocity decreases as a function of the polar angle $\theta$. The red line indicates the position of the disk equator. \label{velFrac}}
\end{figure}

A first analysis of the molecular emission in the ALMA data of L$_{\rm 2}$ Pup by \citet{Kervella2016} has shown that the differential rotation of the equatorial regions of the inner gas disk (within a radius of 6 AU) follows Keplerian rotation. This analysis was primarily based on the $^{\rm 29}$SiO emission, which has the highest signal-to-noise value of the data set. The tangential velocities of the inner disk could be used to estimate the central mass of the system, which was found to be 0.66 $\pm$ 0.05 $\mso$. Beyond this Keplerian region (in the outer disk, i.e. for radii greater than 6 AU) the velocity field  sharply transitions to a sub-Keplerian regime, where the radial velocity decrease is proportional to $r^{-0.85}$. The radius at which the tangential velocity transitions from Keplerian to sub-Keplerian seems to correspond to the inner rim of the dust disk, as determined by \citet{Kervella2015}, indicating that the dust disk may play an important kinematical role in the steady-state hydrodynamical disk system. Following these findings, we separated our velocity model into two regimes, the inner disk ($r < 6\rm AU$), and the outer disk ($r > 6\rm AU$).

\subsubsection{Inner disk}

The tangential gas speeds follow Keplerian rotation:
\begin{equation}
 v_{r_{xy},\rm inner} = \sqrt{\frac{G M_*}{r_{xy}}},
\end{equation}
where $r_{xy}=\sqrt{x^2+y^2}$ is the cylindrical radial distance coordinate, $G$ is the universal gravitational constant, and $M_*$ is the mass of the central object.

We approximated the vertical velocity profile (Fig. \ref{velOff}) in the inner disk. We define the polar angle as
\begin{equation}
 \theta (x,y,z) = {\rm arccos}\left(\frac{z}{\sqrt{x^2+y^2+z^2}}\right),
\end{equation}
assuming the coordinates $(x,y,z)$ make up an ortho-normal set of axes, and the equatorial plane of the disk coincides with the $(x,y)$ plane. To reproduce the vertical velocity structure of the inner disk (Fig. \ref{velOff}), the equatorial tangential velocity field of the inner disk is multiplied with a factor $\zeta$, expressed as
\begin{equation}
 \zeta = (1-f_1)+\left(\frac{f_1}{e^{s_1(\theta-(\pi-\delta))}+1}\right)-\left(\frac{f_1}{e^{s_1(\theta-\delta)}+1}\right),
\end{equation}
where $f_1$ is the factor by which the velocity drops vertically, $s_1$ is the slope steepness of the drop, and $\delta$ is the polar angle at which the drop occurs. The best-fit parameters are summarised in Table \ref{az_par}. This function $\zeta$ is visually represented in Fig. \ref{velFrac}. We hypothesise about the physical origin for the $\zeta$ function in the Sect. 6.1.2.

\begin{table}[htp!]
\centering
\caption{Best-fit vertical velocity profile parameters for the inner disk. \label{az_par}}
\begin{tabular}{ l  l }
\hline
\hline

\multicolumn{2}{ c }{vertical velocity parameters, inner disk} \\
\hline
Parameter & Value \\
\hline
$f_1$ & $0.45$ \\
$s_1$ & $11$ \\
$\delta$ & $\pi/3$ \\
\hline

\end{tabular}
\end{table}

\subsubsection{Outer disk}

A second (sub-Keplerian) velocity field for the equatorial tangential velocities in the outer gas disk ($r > 6\rm AU$) is also based on the preliminary findings of \citet{Kervella2016} and is expressed as
\begin{equation}
 v_{r_{xy},\rm outer} = \sqrt{\frac{G M_*}{6 {\rm AU}}}\left(\frac{6{\rm AU}}{r_{xy}}\right)^{0.85}.
\end{equation}

We found that introducing an additional vertical velocity profile in the outer disk better reproduces the outer edges at high spatial offsets of the wide-slit PVD in Fig. \ref{PV2_data}. Eq. 4 can thus be improved by multiplication with a factor $\alpha$, expressed as
\begin{equation}
 \alpha = (1 - f_2) + f_2\ {\rm exp}\left ( \frac{-z^2}{2 s_2 ^2}\right ).
\end{equation}
In this expression $f_2$ represents the percentage factor by which the equatorial tangential velocity is reduced and $s_2$ is a measure for the rate at which the velocity drops to $(1-f_2) v_{r_{xy},\rm outer}$. The parameter values found to best reproduce the emission patterns are listed in Table \ref{az_par2}.

\begin{table}[htp!]
\centering
\caption{Best-fit vertical velocity profile parameters for the outer disk. \label{az_par2}}
\begin{tabular}{ l  l }
\hline
\hline

\multicolumn{2}{ c }{vertical velocity parameters, outer disk} \\
\hline
Parameter & Value \\
\hline
$f_2$ & $0.1$ \\
$s_2$ & $1.5 \rm AU$ \\\hline

\end{tabular}
\end{table}
The $\alpha$ factor simply reduces the tangential velocities of the regions with high vertical offsets by 10$\%$. The continuous transition from the equatorial tangential velocities to the reduced tangential velocities follows a Gaussian trend, with a characteristic width of 1.5 AU. This means that the tangential velocities are fully reduced to 90$\%$ of their equatorial values at a vertical distance of around 3 AU from the equator.

\subsubsection{Small-scale random motions}

We used the emission gap around the systemic velocity to determine the velocity field of the small-scale random motion (SSRM). Assuming the gap is indeed created by the absorption of the $^{\rm 12}$CO emission by the cold outer regions of the disk, the width of the gap can be used as a proxy for the SSRM. The emission gap spans a width of $1.7 \kms$. Assuming such motions to follow a Gaussian distribution, such a gap thus corresponds to a SSRM standard deviation of $v_{\rm ssrm} = 0.55 \kms$. Because we have no means of probing the SSRM for the regions closer to the star, we assume it to be constant throughout space. However, one can expect the SSRM to increase more deeply inside the disk. We discuss this further in Sect. 6.1.2.

\subsection{H$_{\rm 2}$ density constraints}

The morphology of the disk is properly captured by assuming the following expression:
\begin{equation}
 \rho(r_{xy},\phi,z) = \rho_0\left(\frac{r_{xy}}{r_{\rm c}}\right)^{p}\exp\left[\frac{-z^2}{2H(r_{xy})^2}\right], 
\end{equation}
with $r_{\rm c}$ the inner rim of the gas disk (in the equatorial plane), $\rho_0$ the H$_{\rm 2}$ gas density at $r_{\rm c}$, $p$ the rate with which the density subsides radially, and $H(r_{xy})$ the Gaussian scale height of the disk, expressed as
\begin{equation}
 H(r_{xy}) = H_c\left(\frac{r_{xy}}{r_{\rm c}}\right)^h,
\end{equation}
where $H_c$ is the initial scale height at the location of the inner rim in the midplane of the disk, and $h$ a measure for the rate at which the scale height increases. The radial distance variable $r_{xy}$ is expressed as the cylindrical radial coordinate $\sqrt{x^2+y^2}$. This density expression follows from the solution of a thin, vertically isothermal, non-self-gravitating disk in hydrostatic equilibrium. Although the data indicate that the density profile is more complex, this simplified expression for the profile is sufficiently accurate for the modelling.  We reflect upon this in Sect. 6.3.2.

\begin{table}[htp!]
\centering
\caption{The model density parameter values of the obtained best fit. \label{dens_par}}
\begin{tabular}{ l  l }
\hline
\hline

\multicolumn{2}{ c }{Model density parameters} \\
\hline
Parameter & Value \\
\hline
$r_{\rm c}$ & $2.0 {\rm\ AU}$ \\
$H_c$ & $1.5 {\rm\ AU}$ \\
$h$ & $0.20$ \\
$\rho_0$ & $9.3 \times 10^{-10} {\rm\ kg/m}^3$ \\
$p$ & $-3.1$ \\ 
\hline

\end{tabular}
\end{table}

\citet{Kervella2016} have shown that the gas in the inner disk (< 6 AU) is undergoing Keplerian rotation. This allows us to make an estimate of the radius of the inner rim using the largest observed velocities in the disk signal. The data shows the faintest beginnings of a disk signal approximately 17 $\kms$ away from the central velocity. Assuming a central mass of 0.659 $ \mso$ \citep{Kervella2016}, this tangential velocity translates to the inner rim of the gas being located at approximately $r_{\rm c} = 2.0 \rm AU$.

The evolution of the height of the gas disk as a function of radial distance in the disk plane could be readily deduced from the shape of the disk emission in the low-velocity channel maps (Fig. \ref{chan_contsub}). They show that the disk has a degree of flaring. However, contrary to the archetypal appearance of a flaring disk, the rate of flaring of the disk of L$_{\rm 2}$ Pup decreases as a function of distance from the central star. This trait is reflected in the value of the parameter $h$, which has been determined to be less than unity. Finally, combining the location of the inner rim with the flaring rate, the height of the emission in each channel is best reproduced for an initial scale height of 1.5 AU.

The radial behaviour of the density is captured in the parameters $\rho_0$ and $p$. The value of $\rho_0$ is empirically determined by the iterative efforts targeting the convergence of model versus $^{\rm 12}$CO data emission values in the innermost regions of the disk. The parameter $p$ determines the rate at which the radial density drops, but this parameter cannot be determined from modelling the $^{\rm 12}$CO emission. This is due to the overall optically thick nature of the disk, which results in a substantial insensitivity of the radiative transfer model to the parameter $p$. The optically thin $^{\rm 13}$CO emission permits the evaluation of this parameter $p$. An extensive exploration of the $p$ parameter space has revealed that the data is best modelled by simply assuming a radial power law drop characterised by $p=-3.1$.

The best-fit parameters, which follow from the above-mentioned considerations, are listed in Table \ref{dens_par}.

\subsection{Disk temperature}

The $^{\rm 12}$CO disk must also be optically thick in its more inner regions because it is optically thick at the disk's outer edges. Thus, the $^{\rm 12}$CO emission patterns are mainly produced on the surface boundary confining the gas emitting at a certain frequency along the line of sight. The emission produced by the bulk of the gas lies hidden behind this surface. Therefore, the $^{\rm 12}$CO emission probes the temperature of the gas in this surface layer very accurately and is mostly insensitive to the distribution of mass. Hence, because of the extremely high spectral resolution of the data, these surfaces are only spatially separated by a small distance, and the temperature can therefore be accurately and sufficiently sampled all throughout the disk, permitting the full temperature profile to be precisely determined from the $^{\rm 12}$CO emission.

\subsubsection{Radial profile}

As seen in Fig. \ref{chan_contsub}, the $^{\rm 12}$CO emission close to the central star is brightest, indicating relatively high gas temperatures. The emission drops substantially within the first hundred mas, after which it drops much slower, suddenly plummeting as the outer edge of the emission is reached. This suggests a three-staged temperature profile.

\begin{figure}[htp]
\centering
\includegraphics[width=0.4\textwidth]{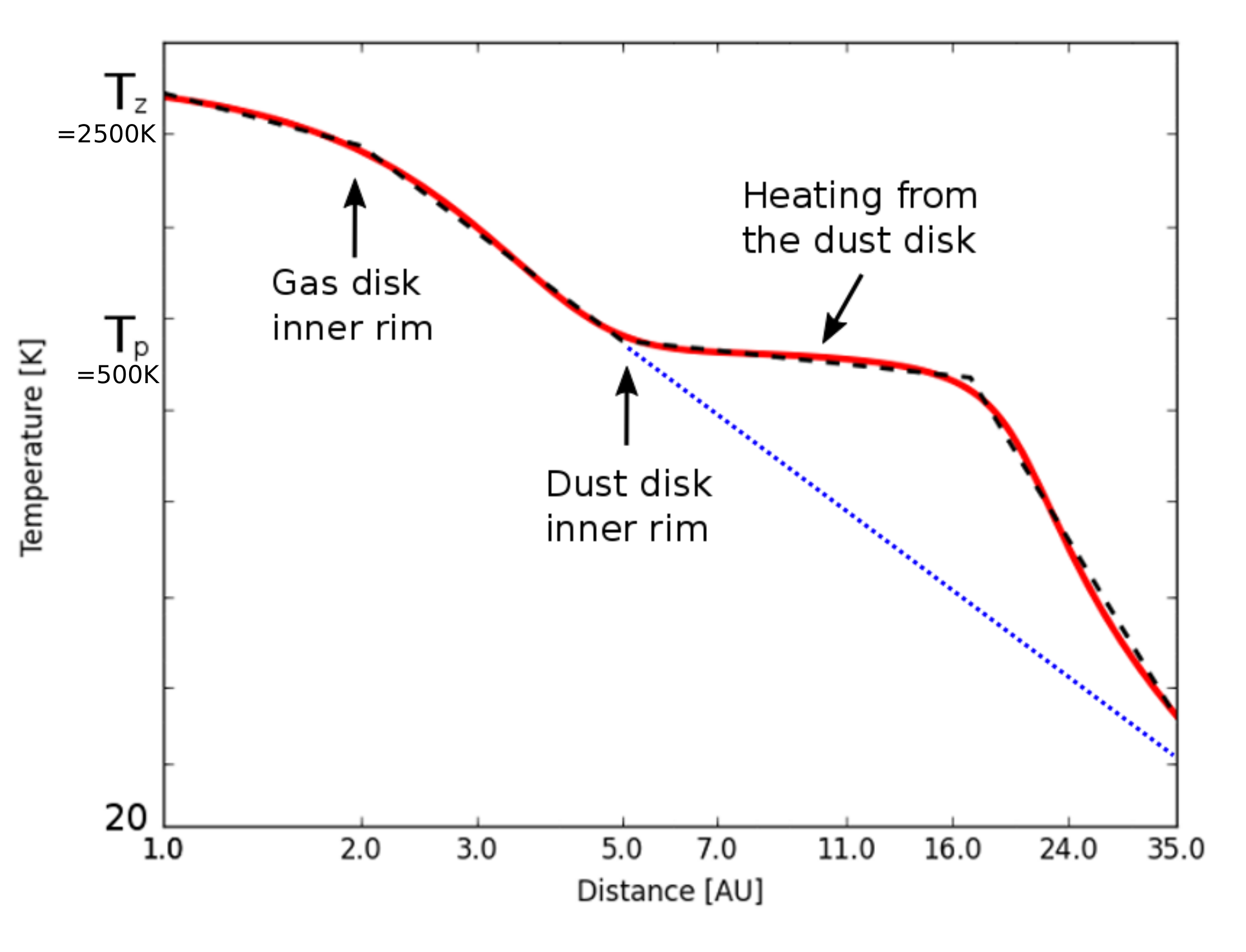}
\caption{Logarithmic plot of the radial temperature profile used to model the $^{\rm 12}$CO disk surrounding L$_{\rm 2}$ Pup, as a function of distance from the central star. The red line represents the continuous expression. The black dotted line is the closest approximation of the red curve by a set of subsequent power laws. The blue dotted line continues the temperature trend of the gas, devoid of dust. The temperature parameter values, and the slopes of the power laws are listed in Table \ref{temp_par}. \label{temp}}
\end{figure}

Analytically, the temperature profile producing the best-fit model is expressed as

\begin{equation}
\begin{aligned}
 T(r) =  (T_{\rm z}-T_{\rm p})\ {\rm exp} & \left ( -\frac{r_{\rm xy}^2}{2w_{1}^2} \right ) \\
         - & \left ( \frac{T_{\rm p}}{\pi} \right ) {\rm arctan} \left ( \frac{r-D}{w_{2}} - \frac{\pi}{2} \right ),
\end{aligned}
\end{equation}
where $T_{\rm z}$ is the temperature at the inner rim of the disk, $T_{\rm p}$ is the temperature of the plateau, $w_1$ is a measurement for the rate of the first temperature decrease, $D$ is the distance at which the second major drop in temperature is located, and $w_2$ is a measure for the rate of the second temperature drop. Table \ref{temp_par} lists the parameter values that were been found to produce the best fit of the data. The temperature profile is shown in logarithmic scales in Fig. \ref{temp}. The temperature levels-off to a near-constant value of 500 K at a distance of approximately 6 AU, which corresponds with the inner rim of the dust disk observed by \citet{Kervella2015}. This indicates that the dust disk plays an important thermodynamical role in the disk system. The temperature of the inner walls of the gas disk (facing the direct radiation of the stellar photosphere) was determined to be around 2500 K. We discuss this further in Sect. 6.2.1 and 6.2.2.

The obtained radial temperature expression can also be approximated by a sequence of power laws, expressed as $T(r)=T_i(r/r_i)^\epsilon$. Here, $T_i$ stands for the temperature at distance $r_i$, and $\epsilon$ is the slope of the power law. Table \ref{temp_par} lists the power law sequence that best fits the continuous model (shown as black dotted lines in Fig. \ref{temp}).

\subsubsection{Vertical profile}

The high angular resolution of the data allows the identification of some structural complexity in the inner disk. As discussed in Sect. 3.2.2, the channel maps seem to exhibit two patches of increased emission in the inner disk both on the blue- and the red-shifted sides. We interpret this signal as the local heating of the walls of the inner disk facing the central star. We modelled this feature as follows:

\begin{equation}
 T_{\rm z} = T_{\rm w} - (T_{\rm w}-T_{\rm eq})\ {\rm exp}\left ( \frac{-z^2}{2\sigma_{T}^2} \right ),
\end{equation}
where $T_{\rm w}$ is the temperature of the star-facing disk wall, $T_{\rm eq}$ is the temperature of the equatorial disk plane, and $\sigma_{T}$ is the Gaussian one-sigma size of the cool equatorial disk plane. In order to reproduce the correct vertical emission contrast in the inner disk we need the temperature of the central disk plane to be approximately at a temperature of 1000 K. Table \ref{temp_par} lists the parameters that were been found to best fit the data.

\begin{table}[htp!]
\centering
\caption{Best-fit model temperature parameter values of the temperature structure. \label{temp_par}}
\begin{tabular}{ l  l }
\hline
\hline

\multicolumn{2}{ c }{Radial temp. parameters} \\
\hline
Parameter & Value \\
\hline
$T_{\rm z}$ & $2500 {\rm\ K}$ \\
$T_p$ & $500.0 {\rm\ K}$ \\
$w_1$ & $1.8 {\rm\ AU}$ \\
$D$ & $20.0 {\rm\ AU}$ \\
$w_2$ & $4.0 {\rm\ AU}$ \\ 
\hline

\multicolumn{2}{ c }{Radial temp. power laws} \\
\hline
Zone & Power $\epsilon$ \\
\hline
$r \leq 2 {\rm\ AU}$ & $-0.5$ \\
$2 {\rm\ AU} < r \leq 6 {\rm\ AU}$ & $-1.4$ \\
$6 {\rm\ AU} < r \leq 17 {\rm\ AU}$ & $-0.2$ \\
$17 {\rm\ AU} < r$ & $-2.9$ \\
\hline

\multicolumn{2}{ c }{Vertical temp. parameters} \\
\hline
Parameter & Value \\
\hline
$T_{\rm w}$ & $2500 \rm K$ \\
$T_{\rm eq}$ & $1000 \rm K$ \\
$\sigma_{T}$ & $0.75 \rm AU$ \\
\hline

\end{tabular}
\end{table}

\subsection{$^{\rm 12}$CO and $^{\rm 13}$CO molecular abundance}

As stated before, molecular abundance and absolute density are degenerate. We opted not to express the $^{\rm 12}$CO abundance as a function of the spatial coordinates, but rather as a constant value. We found no studies that explicitly focus on the molecular abundance of $^{\rm 12}$CO in L$_{\rm 2}$ Pup have been found in the literature. Additionally, the handful of studies describing the general parameters of disks around evolved stars indicate the $^{\rm 12}$CO abundance to be around $10^{-4}$ \citep[e.g.][]{Bujarrabal2013,Bujarrabal2013b}. Studies of $^{\rm 12}$CO abundance values inside protoplanetary disks around young stars also seem to reproduce this value \citep[e.g.][]{France2014}. In addition, because of its high binding energy, carbon monoxide is generally almost chemically unreactive \citep[e.g.][]{Gobrecht2016}.

We have thus chosen to set the $^{\rm 12}$CO/H$_{\rm 2}$ ratio to the constant value of $10^{-4}$ for this model; this value which lies within the reasonable range of expected $^{\rm 12}$CO abundance values for AGB stars \citep[e.g.][]{Decin2010, DeBeck2012}.

The $^{\rm 13}$CO abundance is easily acquired from the values of the physical properties of the inner rim determined from the $^{\rm 12}$CO analysis as boundary condition. The $^{\rm 13}$CO molecular abundance was iteratively adapted until the emission of the inner rim of the $^{\rm 13}$CO model matched the data. This resulted in an abundance estimate of $10^{-5}$, resulting in a $^{\rm 12}$CO over $^{\rm 13}$CO abundance of 10.

\section{Radiative transfer results}

\begin{figure}[htp!]
\centering
\includegraphics[width=0.5\textwidth]{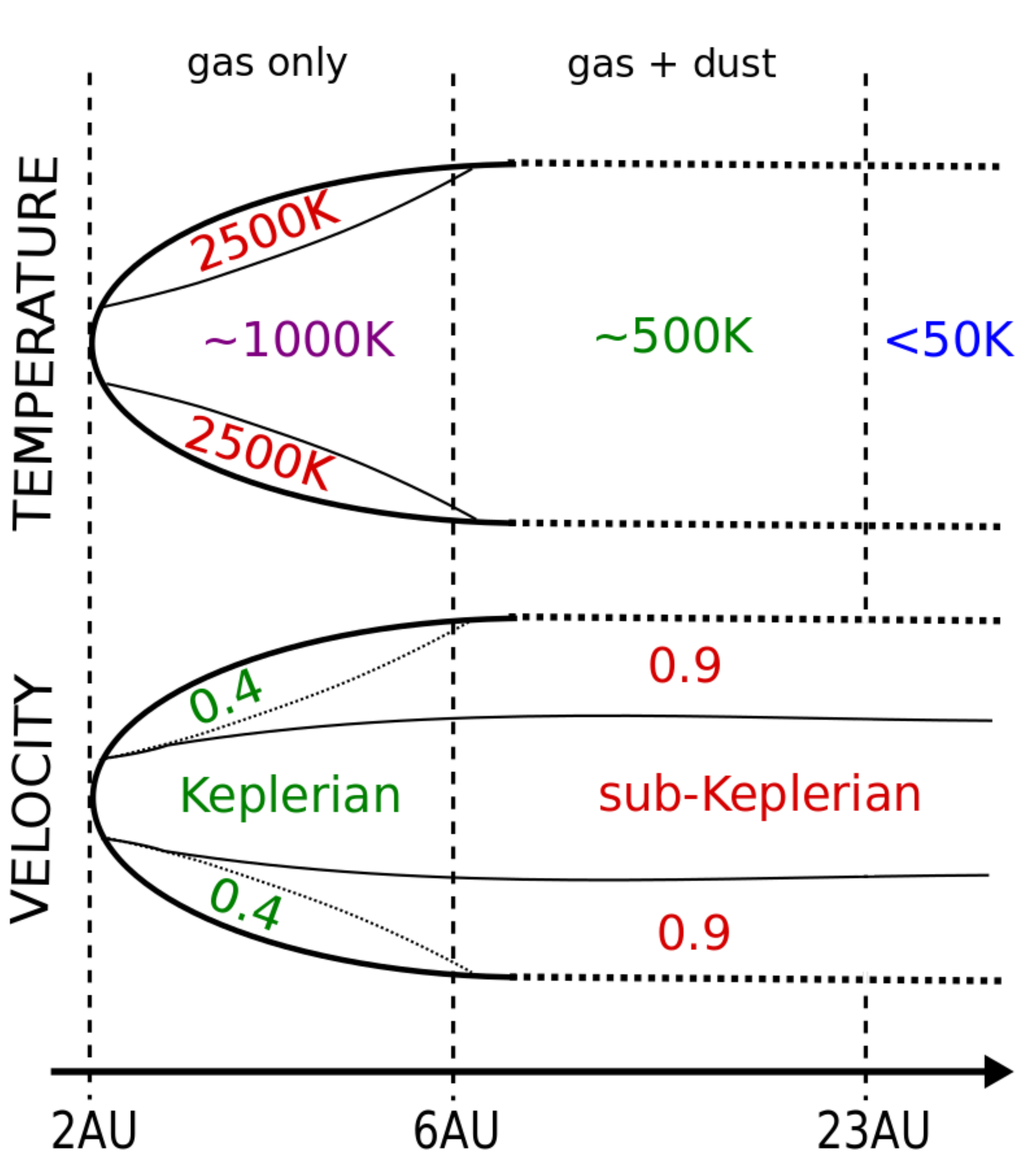}
\caption{Schematic representation of the inner disk velocity and temperature structure, as deduced from the radiative transfer modelling. The global and internal features are not to scale. In the velocity diagram, the numerical values correspond to the multiplication factor with which the equatorial speeds are multiplied to obtain the lowest measured velocities in the regions with high vertical offsets. The colour coding makes the distinction between the inner disk (r < 6 AU) and the outer disk (r > 6 AU). The radial density structure is not represented here, it is simply characterised by a power law with slope $-3.1$. \label{schem}}
\end{figure}

The internal temperature and velocity structure of the disk around L$_{\rm 2}$ Pup is visually represented by the schematic diagram shown in Fig. \ref{schem}. This physical set-up is propagated through the RT code {\tt LIME}, which produces velocity-dependent maps of intrinsic line emission from the species in question. The {\tt CASA} observation simulation scripts transform the intrinsic maps, by introducing observational and instrumental effects, so they can be compared one-to-one with the ALMA data. We simulated the observables for an almost edge-on appearance of the disk, at an inclination of 82 degrees.

These predicted channel maps are represented visually in Figs. \ref{chan_model} and \ref{chan_model2}. Here one can see that the overall morphology and general features present in the ALMA data have been well reproduced. One must keep in mind that all emission in the ALMA data which is smooth on scales greater than 250 mas has been insufficiently sampled in uv space, generating the (seemingly) randomly scattered emission patches that are most strongly present around systemic velocity in the $^{\rm 12}$CO maps. The bulk of this emission probably makes no part of the disk system, and it is therefore not present in the model data.

\begin{figure*}[htp!]
\centering
\includegraphics[width=0.8\textwidth]{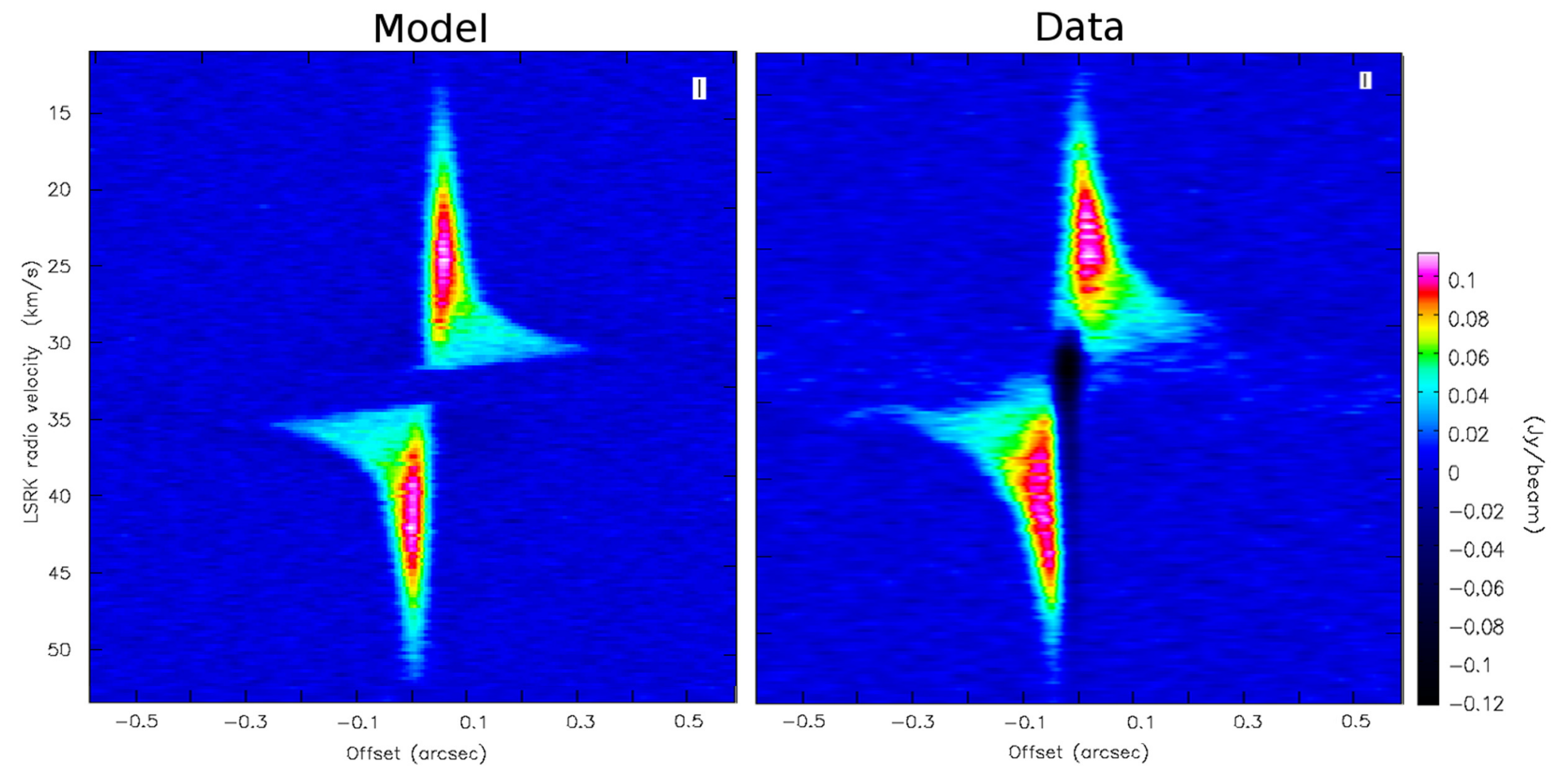}
\caption{Comparison of the $^{\rm 12}$CO PVD constructed by taking a slit along the disk with a minimal width of 15 mas. The left panel shows the PVD of the radiative transfer model, the right panel shows the PVD of the data. $v_{\rm sys} = 33.3 \kms$. \label{PV_1pix_model}}
\end{figure*}

\begin{figure*}[htp!]
\centering
\includegraphics[width=0.8\textwidth]{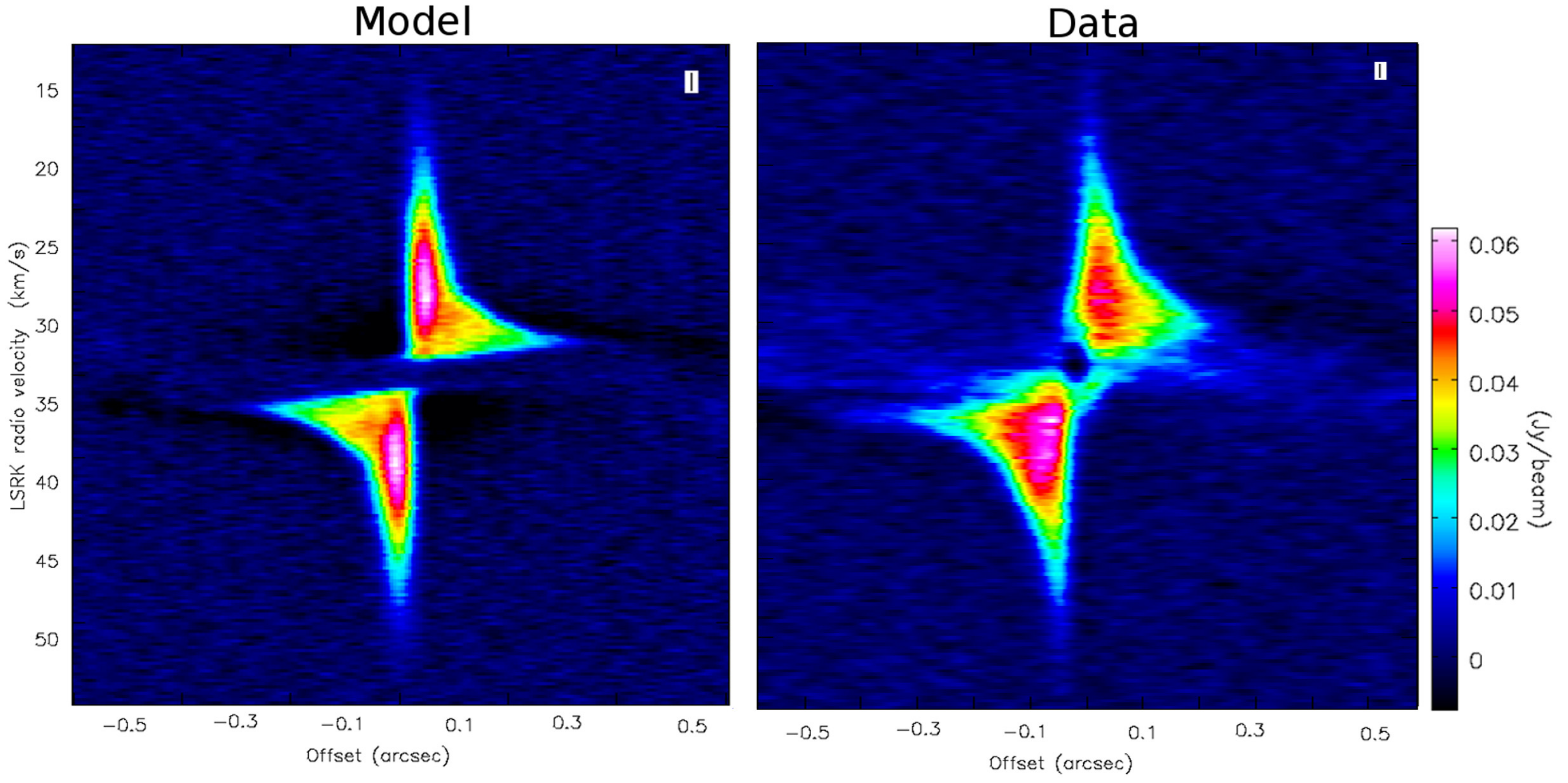}
\caption{Comparison of the $^{\rm 12}$CO PVD constructed by taking a slit along the disk with a width equal to the width of the disk in the data, which is 280 mas. The left panel shows the PVD of the radiative transfer model; the right panel shows the PVD of the data. For reference, $v_{\rm sys} = 33.3 \kms$. \label{PV_99pix_model}}
\end{figure*}

\begin{figure*}[htp!]
\centering
\includegraphics[width=0.8\textwidth]{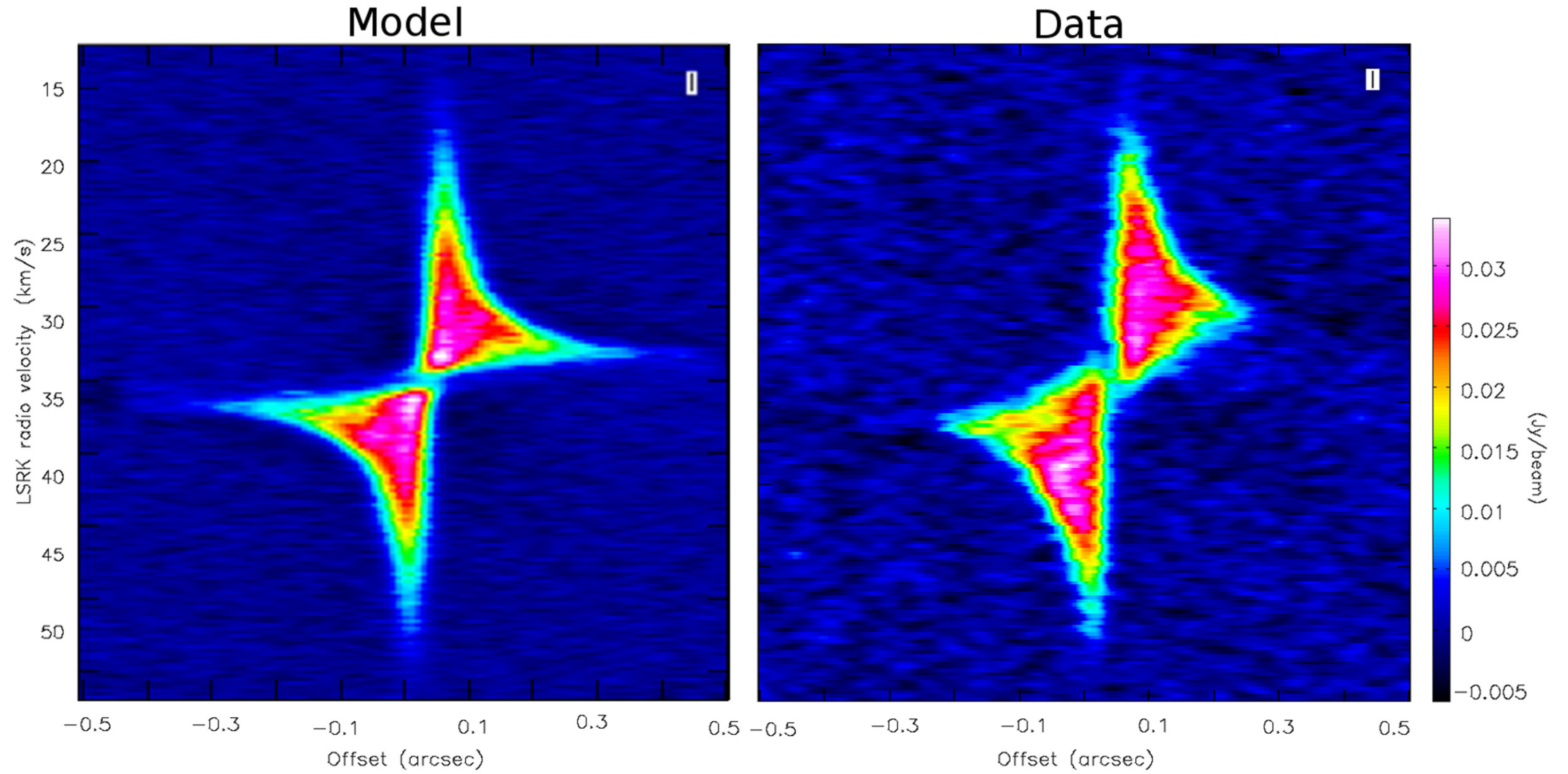}
\caption{Comparison of the $^{\rm 13}$CO PVD constructed by taking a slit along the disk with a width equal to the width of the disk in the data, which is 190 mas. The left panel shows the PVD of the radiative transfer model; the right panel shows the PVD of the data. For reference, $v_{\rm sys} = 33.3 \kms$. \label{PV_99pix_model13}}
\end{figure*}

It is quite difficult to compare the model and data channel maps one-to-one. We shall thus focus the rest of the comparative discussion on the PVDs, which are shown in Figs. \ref{PV_1pix_model}, \ref{PV_99pix_model}, and \ref{PV_99pix_model13}. Fig. \ref{PV_1pix_model} nicely illustrates the quality of the $^{\rm 12}$CO model along zero spatial offset, which corresponds to the disk midplane. All major and minor features present in the data have been reproduced. We succeeded in replicating the edges of the emission signal in velocity space, the absolute velocity magnitude of the offset peaks, the width(in velocity space) of the emission gap around the central velocity and the internal emission distribution of the PVD. The curvature of the signal, which characterises how the outer edge of the disk emission varies as a function of velocity has also been modelled properly. Also the absolute emission values have been accurately reproduced. The major difference is found when comparing the signal near the velocities closest to $v_{\rm sys}$. The outer regions of the disk are probably quite turbulent on different length scales, causing the offset of the lowest velocity regions to decline in a more irregular fashion compared to the analytical model.

The limitations of the model become fully apparent in the full disk-width slit PVDs, shown in Fig. \ref{PV_99pix_model}, where important discrepancies between the data and the model appear. The absolute level of emission around zero spatial offset is slightly overestimated by the model. This is probably caused by the introduction of a vertical temperature dependence to model the temperature difference between the walls of the disk and its equatorial regions. The model most probably overestimates the size of the hottest regions, contributing to the overestimated average emission around zero offset. There is also an emission distribution asymmetry between the east and west wings of the data, which is of unknown origin, but can perhaps be attributed to the presence of a binary companion, as suggested by \citet{Kervella2016}. Close to the central velocities one can clearly identify emission in the data that is not present in the model. We believe this emission originates completely from the poorly reconstructed large-scale emission from the CSE outside the disk, observed by the (poorly covered) shortest baselines of the configuration.

The final PVD comparison we consider is the full disk-width slit PVD of the $^{\rm 13}$CO data, shown in Fig. \ref{PV_99pix_model13}. We do not focus on the narrow-slit PVD of the $^{\rm 13}$CO emission as it is virtually identical to the discussion on the $^{\rm 12}$CO narrow-slit diagram. The emission distribution of the wide-slit PVD has been well reproduced by the model, and so has the absolute level of the emission. However, a clear difference is seen in the extent and location of the maximum spatial offset peaks. The peaks of the model are very pointy and long, and are not as widely separated in velocity space as the offset peaks in the data. We believe this difference can be attributed to photodissociation effects, which could affect the outermost regions of the $^{\rm 13}$CO disk. Accounting for this effect could thus seriously reduce the maximal spatial extent of the offset peaks in the model PVD. In addition, the location of the peaks in velocity space would then also shift to higher projected velocities. 

\section{Discussion}

\subsection{Parameter sensitivity}

Because we cannot perform an extensive and in-depth exploration of the parameter space available to us (see Sect. 3.3), we cannot quantify the quality of the model in terms of statistically significant errors. Hence, we dedicate this section to a discussion of the uncertainties in the estimated parameter values.

The kinematical properties of the disk can be directly measured from the data. We are thus confident that both the equatorial and vertical velocity profile are accurately constrained and the parameter values that have been deduced directly from the velocity field can be trusted. Furthermore, the geometrical properties of the system can also be directly extracted from the data. We do not expect major uncertainties here either.

Carbon monoxide has a very low electric dipole moment causing the density, temperature, and molecular fractional abundance (DTA) parameters to be completely degenerate. Fortunately, the modelled data consisted of both optically thick and thin emission. In case of the optically thin $^{\rm 13}$CO lines, one cannot pin down the exact combination of temperature and density properties of the disk and the abundance of this molecule. In case of the optically thick $^{\rm 12}$ lines the degeneracy of temperature and density profile is mostly lifted, and the temperature profile can be determined. However, the assumed value for the $^{12}$CO molecular abundance remains an uncertainty.  The adopted value, from reasonable and justifiable arguments, directly and linearly affects the disk density. For example, an increase in the molecular abundance by a factor 2 should result in a decrease in density by the same factor. Hence, the largest uncetainty relates to the determined value for $\rho_0$. But, as argued in Sect. 4.4, we do not expect the true abundance fraction to deviate substantially from the assumed one. The radial density power law should not be affected by the abundance estimate, as it is mostly dependent on the assumed temperature profile.

The optical thickness of the $^{\rm 12}$CO emission, and the high spectral resolution of the data, has permitted us to very accurately constrain the equatorial temperature profile. And assuming the adopted $^{\rm 12}$CO abundance fraction is correct, then the retrieved equatorial density parameters and $^{\rm 13}$CO abundance value are also very tightly constrained. However, where the $^{\rm 12}$CO emission becomes optically thin, the subsequently constrained parameter values likely have much larger uncertainties. In the inner disk, the low density regions of high vertical offset are cut off by the shearing of the stellar outflow. Here the $^{\rm 12}$CO emission does not become optically thin, and the free parameters are thus expected to be tightly constrained. At the top and bottom edges of the outer disk, the $^{\rm 12}$CO emission definitely turns optically thin. This locally reinstates the full DTA degeneracy, therefore breaking down the assumption that the temperature can be directly measured from the emission. However, the regions in question are small compared to the full disk size. Thus, increased local uncertainties on the derived parameter values probably does not affect the final result by too great an amount.

\subsection{Velocity field}

\subsubsection{Sub-Keplerian regime of the outer disk}

The sub-Keplerian regime of the disk has a power law dependence of the tangential velocity that is proportional to $r^{-0.85}$ \citep{Kervella2016}. This suggests the presence of a mechanism that slows down the gas and whose effectiveness increases with radius. One explanation for such a deviation from Keplerian rotation is a radial force imbalance. However, in this scenario the sub-Keplerian nature of the outer disk should result in the appearance of an inward motion of the gravitationally attracted material. Such an inward velocity component has not been detected in the emission (see \citet{Homan2016} for an overview of the expected emission patterns caused by such an additional inward velocity component). This means that either the projected inward velocity component is negligibly small compared to the spectral resolution, or a mechanism exists that generates an additional outward force which balances the gravitational pull, yet increases in strength with distance. A possible origin for such a force would be the radial stratification of dust grain sizes within the disk. Such a dependence, with larger grains closer to the star, can produce a radial radiation pressure profile that is able to alter the effective gravity felt by the dust grains throughout the disk (see upcoming paper). Assuming a tight mechanical coupling between dust and gas, this scenario should then act as a cumulative reduction of the effective gravity felt by the surrounding gas as a function of radial distance.

\subsubsection{Vertical velocity profile of the inner disk: The $\zeta$ and $\alpha$ functions.}

We hypothesise that the $\zeta$ profile (Fig. \ref{velFrac}), detailing the vertical distribution of tangential velocities in the inner disk, originates from hydrodynamical boundary instabilities at the disk/radial outflow interface. Media subjected to velocity shear will develop Kelvin-Helmholtz instabilities \citep[KHI; ][]{Funada2001}, which destroy any original organised velocity field, causing turbulent features of high complexity to arise. For a physical set-up like the inner disk of L$_{\rm 2}$ Pup, the timescale for KHI to arise is of the order of years \citep{Funada2001}. The unstable regions of the system will converge to a maximum entropy state via mixing when left to operate for a long time (about $10^3 \times$ KHI timescale). For the L$_{\rm 2}$ Pup system this would be of the order of thousands of years, which is much shorter than the AGB lifetime. The resulting velocity field inside this mixed region is an averaged distribution over all velocities that were part of the interaction. Thus, a radial outflow (with no tangential velocity component) shearing against a rotating gas disk creates KHI, which average out to an effective decreased rotation speed. Similarly, one can expect a small radial velocity component to be present in this 'boundary region'. However, because the radial component would be small, and because of the orientation of the radial field with respect to the observer, the effect would potentially only be visible in the channel maps close to central velocity. The emission in these channel maps is hidden behind the optically thick, cold outer disk edge. Thus, we are unable to confirm the presence of this small radial velocity component in the regions with a high vertical offset of the inner disk.

The origins of the $\alpha$ profile, which modifies the vertical profile of the rotational velocity in the outer disk, can be attributed to the same effect. However, because the relative velocities between the disk and the outflow are smaller in the outer disk compared to the inner disk, the magnitude of the effect is reduced.

\subsubsection{Velocity field of the small-scale random motion}

\begin{figure}[htp!]
\centering
\includegraphics[width=0.5\textwidth]{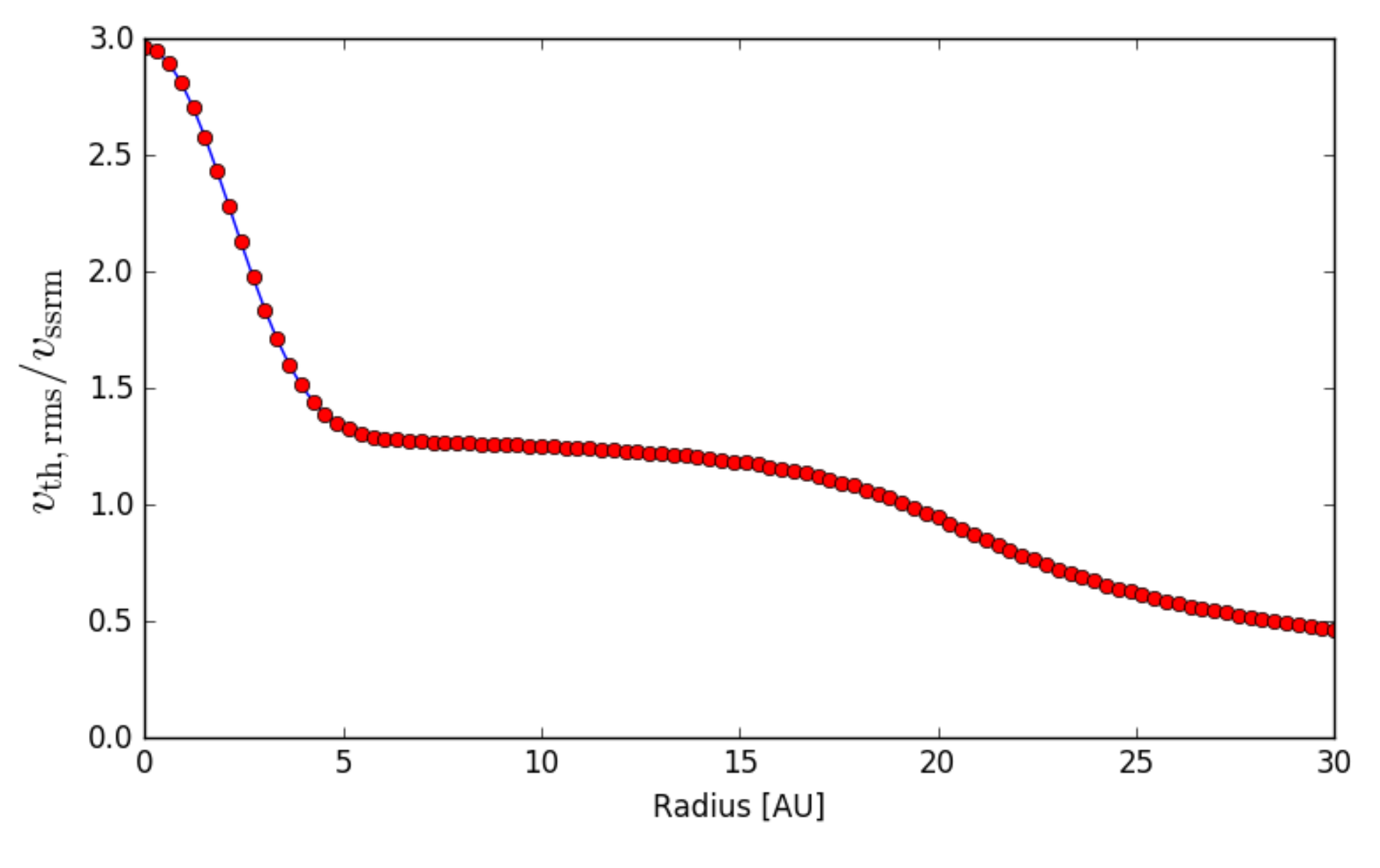}
\caption{Rms thermal velocity (as derived from the temperature profile) relative to the assumed turbulent velocity as a function of radius. \label{ehsan}}
\end{figure}

We derived the magnitude of $v_{\rm ssrm}$ from the (lack of) spectral signatures of the gas located in the outermost regions of the disk. We do not possess a complete description of how one can expect the SSRM of a medium to depend on the physical properties of that medium. Therefore, to limit the number of assumptions made, we approximated the SSRM as constant throughout the disk. However, this assumption is rather unlikely. The rms thermal speed of molecules at a certain temperature is given by 
\begin{equation}
 v_{\rm th, rms} = \sqrt{\frac{3 k_B T}{m}},
\end{equation}
where $k_B$ is the Boltzmann constant, $T$ is the temperature of the gas, and $m$ is the molecular weight. The rms thermal speed reaches values up to 1.5$\kms$ for CO molecules at 2500 K, which is significantly greater than the $v_{\rm ssrm} =\ $0.5$\kms$ value used in this paper. In addition, as can be inferred from Fig. \ref{ehsan}, the thermal velocity in the cool outer regions only makes up about 50$\%$ of the deduced SSRM. This is a strong indication that the small-scale velocity in the disk consists of additional components besides the thermal rms speed and that our assumed $v_{\rm ssrm}$ probably underestimates its true physical value inside the disk. 

\subsection{Temperature structure}

\subsubsection{Vertical temperature profile of the inner disk}

The modelling strongly suggests the requirement of a vertical temperature profile describing the extent by which the inner disk walls facing the direct stellar radiation field are heated. The maximum temperature of the gas in these regions was determined to be approximately 2500 K. This can be readily explained by thermal coupling of the gas with the stellar radiation field. The disk walls directly facing the stellar radiation heat up to high temperatures, while shielding the equatorial regions of the disk, keeping them at cooler temperatures.

\subsubsection{Radial temperature profile}

The shape of the radial temperature profile (Fig. \ref{temp}) seems rather peculiar. However, the regime transition between the inner disk and the outer disk seems to strongly indicate that the presence of dust plays an important thermodynamical role. We hypothesise the following explanation. In the inner disk, the temperature profile is simply set by the dissipation of heat, provided by the photospheric resonance photons. From 6 AU onward, dust is prominently present in the disk. Thus, from this radius outward a substantial increase in sensitivity to the photospheric continuum emission exists. Assuming a tight mechanical coupling between dust and gas (which the velocity profile suggests), the dust provides a continuous source of heat from inside the disk. This lessens the radial rate of temperature decrease and acts as a thermal buffer. This happens for as long as an interaction between the stellar radiation field and the dust in the disk occurs. However, finally, at a specific radius one can expect the dust to become fully optically thick (radially), prohibiting the further propagation of energy via photospheric continuum photons, causing a final temperature drop at large radii.

\subsection{Density structure}

\subsubsection{Disk mass}

From the density structure of the disk we may calculate its mass. Integrating over an infinite volume the mass of the disk is found to be approximately $2.2 \times 10^{-4} \mso$, or about 0.23 Jupiter masses. We analysed the sensitivity of this value to the determined density parameters. Assuming each parameter would have its true physical value between plus and minus 25\% of the determined value, then the disk mass would lie between $0.65 \times 10^{-4} \mso$ and $9.7 \times 10^{-4} \mso$. The determined disk mass is most sensitive to the exponent of the radial power law. If we assume the same spread on the parameter values as before, but now keep the radial power fixed, the spread on the disk mass diminishes significantly. It would lie between $1.15 \times 10^{-4} \mso$ and $3.4 \times 10^{-4} \mso$. These values are in agreement with the estimates of \citet{Kervella2014,Kervella2015}.

\subsubsection{Density profile}

As stated in Sect. 4.3, the density description conforms with the solution of a thin, vertically isothermal, non-self-gravitating disk in hydrostatic equilibrium. Although on first sight the disk does not seem to be congruent with these assumptions, a deeper consideration of the deduced disk physics can prove otherwise. The channel maps (Figs. \ref{chan_contsub} and \ref{chan_contsub2}) exhibit a rather broad disk with a height of around 300 mas. We find that the observations are well fitted by a disk with a Gaussian scale height (Sect. 4.3.2), which means that the bulk of the disk is in fact located in a thinner slab around the equator. Around 75\% of the total disk material is located in a 5 AU (78 mas) wide slab. The mass of the disk is approximately 1.5\% of the central mass, making its self-gravity negligible with respect to the primary gravitational source. The largest part of the disk has been found to be vertically isothermal. This property has only been found to break down in the star-facing wall of the inner disk. The resultant features, however, are only present in the thin interaction region at the disk-outflow interface. The effect of this local heating on the overall thermal disk structure is negligible. In fact, this vertical temperature profile of the inner disk could be the cause for the $h=0.2$ radial scale height growth to be lower than expected (< 1). As for vertical hydrostatic equilibrium (VHE), this is very difficult to deduce from the observations. One can only assume that if the disk is indeed a stable structure present around the AGB star, then VHE must be satisfied. It therefore seems that the assumptions contained within the analytical density description may very well be satisfied.

The value $p=-3.1$, which characterises the slope of the radial density power law, seems very steep compared to the typical slopes presented in the literature. Typical density slope values for disks around post-AGB stars and planetary nebulae are of the order of $p=-2.0$ \citep[e.g.][]{Bujarrabal2013b, Bujarrabal2015, Bujarrabal2016}. Protoplanetary disks seem to possess even flatter profiles with $-2.0 \leq p \leq -1.0$ \citep[e.g.][]{Hartmann1998,Zhang2014,Podio2014,Kirchschlager2016}. Rings of asteroids in dynamic resonance with a planet, such as in the asteroid belt of our solar system, have an even lower value of $p \sim -0.1$ \citep{Ryan2009}.

\subsection{Evolutionary considerations}

\subsubsection{Disk origin}

The scenario that the disk is a remnant from an earlier evolutionary phase is improbable as protoplanetary disks disperse fairly early on in the main sequence phase \citep[e.g.][]{Kimura2016}. Therefore, the disk must have been formed during the later evolutionary stages of the star. A scenario for a recent formation of a disk may be that of a latitude dependent mass loss in the AGB phase. However, it seems highly unlikely that such a mechanism produces a disk system that is as vertically compact as observed here. Moreover, it needs a mechanism to prevent the material escaping from the system.

One of the most plausible scenarios for the formation of an equatorial density enhancement in the CSE of an AGB star is by means of a binary companion \citep{Theuns1993,Soker1994,Mastrodemos1999,Kim2011,Mohamed2012}. Such a companion funnels part of the wind material in its gravitational field, expelling it as a compact stream along the orbital plane. \citet{Kervella2014,Kervella2015,Kervella2016} have suggested the presence of a binary companion around L$_{\rm 2}$ Pup, based on precise measurements of continuum asymmetries. The fact that the inner rim of the gas disk (determined in this work) coincides with the location of the observed candidate companion of L$_{\rm 2}$ Pup in literature supports this hypothesis.

The presence of a binary companion is also supported by angular momentum considerations. From the density distribution and velocity structure we calculate the total angular momentum contained within the gas disk to be
\begin{equation}
 L_{\rm tot} = 3.1 \times 10^{42}\ {\rm m^2 kg/s}.
\end{equation}
A first analysis by \citet{Kervella2016} suggests that the star L$_{\rm 2}$ Pup had an initial mass comparable to the mass of the Sun. Recent work has shown the Sun to contain approximately $10^{41}\ {\rm m^2 kg/s}$ of angular momentum \citep{Iorio2012}. In addition, recent astroseismologic analyses of light curves of solar-type stars indicate only minor differences in internal rotation structure \citep{Benomar2015,Schunker2016}. Thus, making the assumption that the total angular momentum of L$_{\rm 2}$ Pup on the main sequence was comparable to that of the Sun, we find that the disk around L$_{\rm 2}$ Pup contains approximately 30 times the total amount of available angular momentum inside the star. The presence of a binary companion could easily explain this angular momentum discrepancy. By assuming the companion follows Keplerian rotation, and by assuming an angular momentum transfer efficiency of 30\% \citep{Nordhaus2006,Nordhaus2007} from the companion to its surroundings, the total angular momentum in the disk could be explained by the presence of a 1 Jupiter mass companion. Extending the 'sensitivity analysis' on the mass of the disk to this discussion (assuming the same uncertainty range of the determined density parameters and assuming zero uncertainty on the velocity field), the mass of the companion could be expected to lie within the 0.25 to 5.5 Jupiter mass range. This estimation is in agreement with the mass range of the companion proposed by \citet{Kervella2016}. These estimates are lower limits, as one can expect a portion of the main sequence angular momentum to have been lost in the stellar wind during previous evolutionary stages.

In paper I, \citet{Kervella2016} discuss the likelihood of the companion planet to have formed inside the disk. After all, the disk contains enough material to build one or more planets. However, as an additional consideration, \citet{Pelupessy2013} argue that the formation of planets inside evolved disks is rather unlikely. Timescale considerations seem to support this. \citet{Miller2016} claim that thermally pulsing AGB lifetimes are typically less than one Myr, while typical planet formation timescales are orders of magnitude larger \citep{Alibert2005,Hori2011,Alibert2013}. In addition, at present no mechanism, other than binarity, can explain the presence and formation of differentially rotating disks around evolved stars, besides binarity. Therefore, the planetary binary companion was probably present around the star before the onset of the AGB.

%We calculate how long it would take the star to produce the mass contained within the disk. Following \citep{Reimers1975},vusing the scaling factor $\eta = 0.477$ \citet{McDonald2015}, we find a mass loss rate for L$_{\rm 2}$ Pup of $6.4 \times 10^{-8} \msoy$. Assuming the disk occupies a 'rigid' scale height of 9 degrees, it would obstruct approximately 15\% of the area of a sphere. Thus, the disk accumulates $\sim 10^{-8}$ solar masses per year of material. Hence, it should take $\sim$22000 years to build up a disk like this, which is much less than the tp-AGB timescale. The disk could thus have formed from the AGB outflow.

Assuming the companion was accompanying the star throughout its life, the disk could either be made up of mainly AGB wind material, or could contain a substantial portion of evaporated planet material. The evolution of the star to the giant phase would have substantially heated the planet, possibly even causing evaporation and outgassing. Using our determined $^{\rm 12}$CO over $^{\rm 13}$CO ratio as a proxy for the $^{\rm 12}$C / $^{\rm 13}$C abundance fraction, we find the first scenario more probable. A $^{\rm 12}$C over $^{\rm 13}$C ratio of 10 (see Sect. 4.4) is typical of oxygen-rich AGB gas. In comparison with typical values in our Sun and solar neighbourhood  \citep{Asplund2009,Kobayashi2011} we would expect a ratio that is an order of magnitude higher were the disk made of substantial amounts of material from planetary origins. This also supports the findings in paper I, where the spectral signatures at the companion location seem to indicate that it is accreting material via wind Roche lobe overflow.

\subsubsection{Further evolution}

\citet{Pelupessy2013} argue that the formation of planets inside evolved disks is unlikely. Only a very narrow combination of physical parameters permit planet formation. The physical properties of the disk of L$_{\rm 2}$ Pup lie well outside this range. Although no new planets are formed in the disk, the question remains whether the ones present can survive. The discovery of planets around a subdwarf-B star \citep{Silvotti2014} shows that planets have survived the tumultuous giant phase. In addition, \citet{DeMarco2015} deduces that some PN in the \emph{Kepler} spacecraft field of view may also contain orbiting planets. Evidence thus strongly suggests that planets can persist through the final stages and influence the circumstellar morphology up to the white-dwarf stage.

Detailed studies of post-AGB morphologies \citep{Bujarrabal2016} show a vast difference in length scale compared to the L$_{\rm 2}$ Pup system with disks ranging up to 1000 AU. In addition, their internal density structure is vastly different from the $p=-3.1$ slope found in this work (Sect. 6.3.2). Not many objects have been studied to a high degree of detail so it is difficult to gauge the significance of the difference between the evolutionary stages. We hypothesise nonetheless, and discuss a few scenarios which could explain the difference.

We speculate that the substantial difference in density distribution may be a consequence of the mass-loss rate of the central star. Low mass-loss rate AGB stars do not provide much material for the planet to interact with. This can be expected to result in a compact and low-density disk with a steep density slope. High mass-loss rates would pump more material into the interaction zone, causing wider and denser disks with flatter density slopes. However, as AGB stars evolve towards the post-AGB phase, it is believed that the mass-loss rate increases. If the circumstellar disk is not dense and massive enough to fully counter weigh this increased outward force, the disk is pushed outwards, slowing down its rotation substantially. This could explain the presence of the giant non-rotating tori observed around many post-AGB stars and planetary nebulae \citep[e.g.][]{Skinner1998,Sahai2006,Sahai2011,Lagadec2011}. Rotating disks that are massive enough to survive this 'super wind' phase remain as such \citep[e.g.][]{Hillen2015,Hillen2016}.

The scale differences between AGB and post-AGB disks could also be explained by morphological constraints. As the star transitions to the AGB, equatorial wind material gets injected into a rotating disk by interaction with an orbiting planet. As more material accumulates, the disk becomes denser. At a certain point in time the density of the disk would reach a point where it would substantially obstruct the movement of the AGB wind along the equator. This would be the current stage of the L$_{\rm 2}$ Pup system, as there is a clear lack of radial gas flow inside the disk. The outflow material left unchanneled by the companion would therefore be forced towards the polar directions (along the path of least resistance), forming a wide bipolar outflow. \citet{Akashi2008} suggest a mechanism via which the disk can accumulate mass expelled by this outflow, which would cause the disk to grow, and the density slope to flatten.

In any case, the study of the CSE of L$_{\rm 2}$ Pup strongly suggests that the presence of low mass (planetary) companions play a vital role in the shaping of the CSE of the AGB star, whose properties dictate the further morphological evolution of the system.

\section{Summary}

The circumstellar environment of L$_{\rm 2}$ Pup was imaged by ALMA at $15 \times 18 \rm\ mas$ resolution. The $^{\rm 12}$CO $J=$3$-$2 and $^{\rm 13}$CO $J=$3$-$2 emission show a rotating gas disk in a nearly edge-on orientation. Using the radiative transfer code {\tt LIME} and the {\tt CASA} post-processing tools, we constrained the physical properties of the disk: 
\begin{itemize}
 \item The radial temperature of the disk is characterised by three major regimes. The temperature of the inner disk (< 6AU) is described by a steep drop from around 1000 K to around 500 K. Beyond this radius the temperature levels off at around 500 K, but drops again steeply to very cold temperatures (< 100 K) at the outer edge of the disk (23 AU). Vertically the temperature of the inner disk is stratified, with cooler (1000 K) equatorial regions and rather hot inner disk walls (2500 K). We attribute this temperature stratification to complex hydrodynamical interactions at the outflow-disk boundary, which mixes dust formed at larger radii back into the innermost regions of the disk.
 \item The density of the disk is well described by the solution for a thin, vertically isothermal, non-self-gravitating disk in hydrostatic equilibrium. Its radial decay is expressed by a power law, with a power of -3.1. The flaring rate of the inner disk is quantified by a power of 0.2. Its inner rim is located at around 2 AU, which coincides with the location of the candidate binary companion detected in paper I. This inner rim has a height of approximately 1.5 AU, indicating that it may be puffed up.
 \item The equatorial velocity was determined by \citet{Kervella2016}, showing a sharp transition from Keplerian to sub-Keplerian at a radius of 6 AU. However, there is also an important vertical dependence, which reduces the expected Keplerian velocities of the inner disk gas to approximately 40\% of its expected tangential velocity at high vertical offsets. The Gaussian turbulent velocity field component was deduced to be 0.5 $\kms$.
 \item The molecular abundance of $^{\rm 12}$CO was assumed to be $10^{-4}$. The molecular abundance of $^{\rm 13}$CO was deduced to be $10^{-5}$, resulting in a $^{\rm 12}$CO over $^{\rm 13}$CO abundance ratio of 10. This is in line with values of other oxygen-rich AGB stars in the literature. This suggests that the bulk of the gas in the disk originates from the star itself.
 \item The dust in the outer disk (r > 6 AU) seems to play a pivotal role in the thermodynamical and kinematical structure of the disk.
 \item Constraining the density profile permitted us to deduce the mass of the disk, which was found to be approximately $2 \times 10^{-4} \mso$. The angular momentum in the disk was found to be over 30 times the expected angular momentum contained within the star L$_{\rm 2}$ Pup. We calculate that a binary companion of $\sim$1 Jupiter mass would be sufficient to explain the angular momentum contained in the disk. 
\end{itemize}

% ***********************************************************************************************************************************************
%                                                                 END MAIN BODY
% ***********************************************************************************************************************************************

\begin{acknowledgements}  
This research made use of the SIMBAD and VIZIER databases (CDS, Strasbourg, France) and NASA's Astrophysics Data System. We acknowledge financial support from the ``Programme National de Physique Stellaire'' (PNPS) of CNRS/INSU, France. W.H. acknowledges support from the Fonds voor Wetenschappelijk Onderzoek Vlaanderen (FWO). LD acknowledges support from the ERC consolidator grant 646758 AEROSOL and the FWO Research Project grant G024112N. This paper makes use of the following ALMA data: ADS/JAO.ALMA\#2015.1.00141.S. ALMA is a partnership of ESO (representing its member states), NSF (USA) and NINS (Japan), together with NRC (Canada) and NSC and ASIAA (Taiwan) and KASI (Republic of Korea), in cooperation with the Republic of Chile. The Joint ALMA Observatory is operated by ESO, AUI/NRAO and NAOJ. KO acknowledges a grant from the Universidad Cat\'{o}lica del Norte. We want to thank the referee for the rigorous review of this manuscript and detailed responses.
\end{acknowledgements}

\bibliographystyle{aa}
\bibliography{wardhoman_biblio}

\IfFileExists{wardhoman_biblio.bbl}{}
 {\typeout{}
  \typeout{******************************************}
  \typeout{** Please run "bibtex \jobname" to obtain}
  \typeout{** the bibliography and then re-run LaTeX}
  \typeout{** twice to fix the references!}
  \typeout{******************************************}
  \typeout{}
 }

\cleardoublepage
\begin{appendix}

\section{$^{\rm 12}$CO channel maps - ALMA data}

\begin{figure*}[h]
\centering
\includegraphics[width=\textwidth]{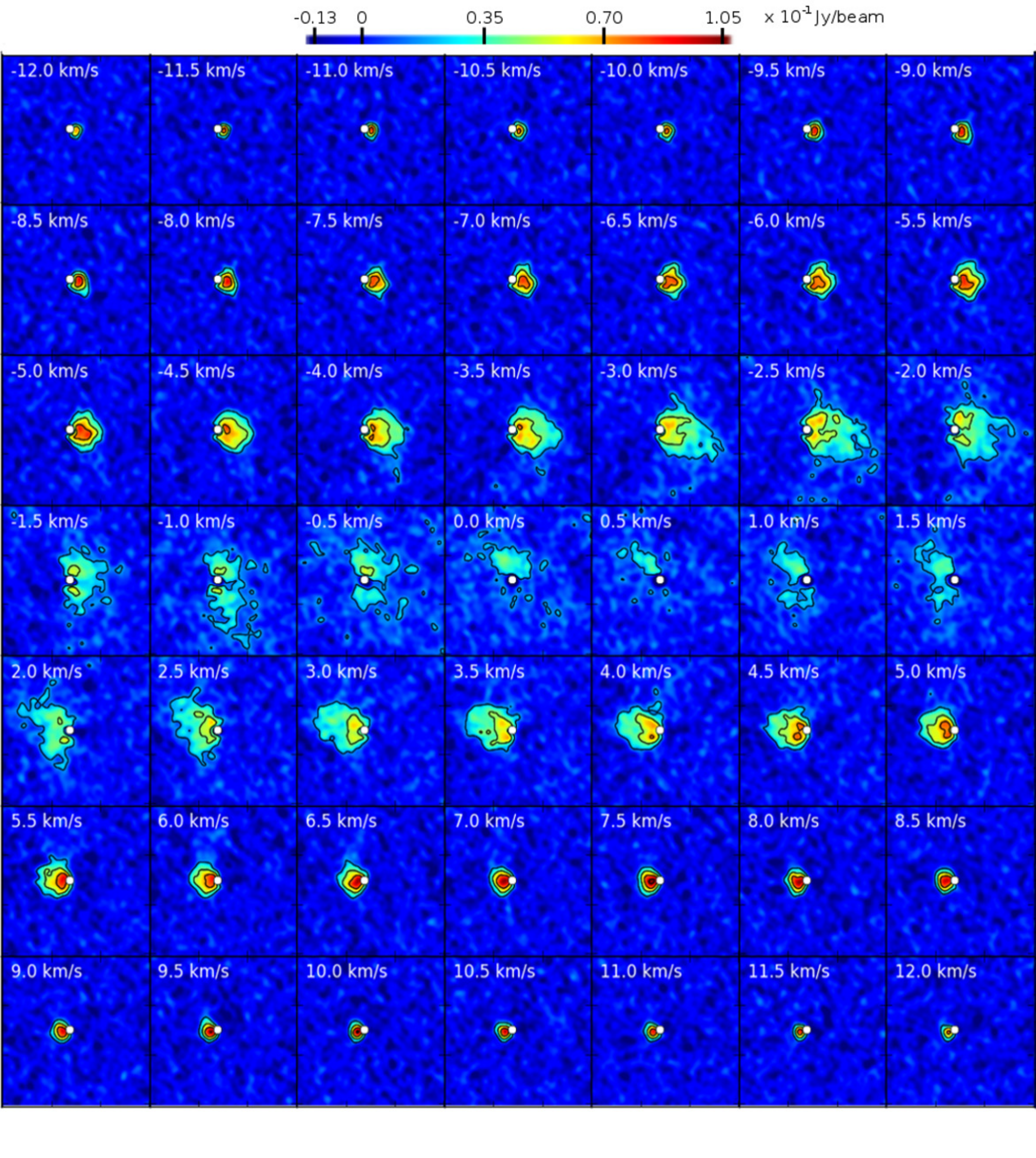}
\caption{Visualisation of the velocity channel maps of the $^{\rm 12}$CO $J=$3$-$2 emission of the circumstellar environment of L$_{\rm 2}$ Pup with the continuum subtracted. Each panel has a dimension of 1'' by 1''. The velocity channel is indicated in the top left corner of each panel. The black contour levels are drawn every 7 times the rms noise value outside the line. The white dot represents the location of the centre of mass. \label{chan_contsub}}
\end{figure*}

\section{$^{\rm 12}$CO channel maps - radiative transfer model}

\begin{figure*}[h]
\centering
\includegraphics[width=\textwidth]{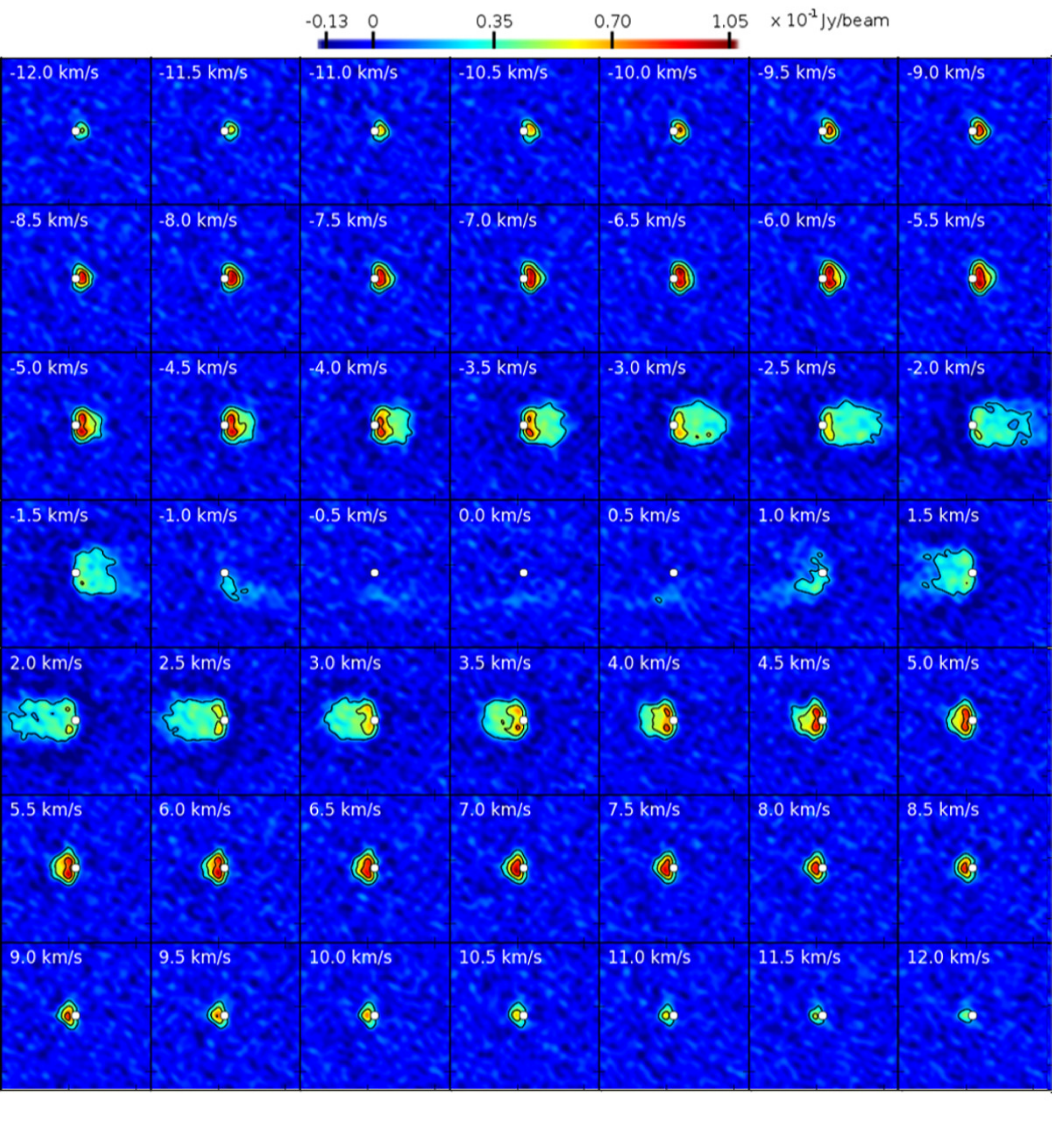}
\caption{Same as Fig. \ref{chan_contsub}, but now showing the synthetic emission of the rotating disk model. \label{chan_model}}
\end{figure*}

\section{$^{\rm 13}$CO channel maps - ALMA data}

\begin{figure*}[h]
\centering
\includegraphics[width=\textwidth]{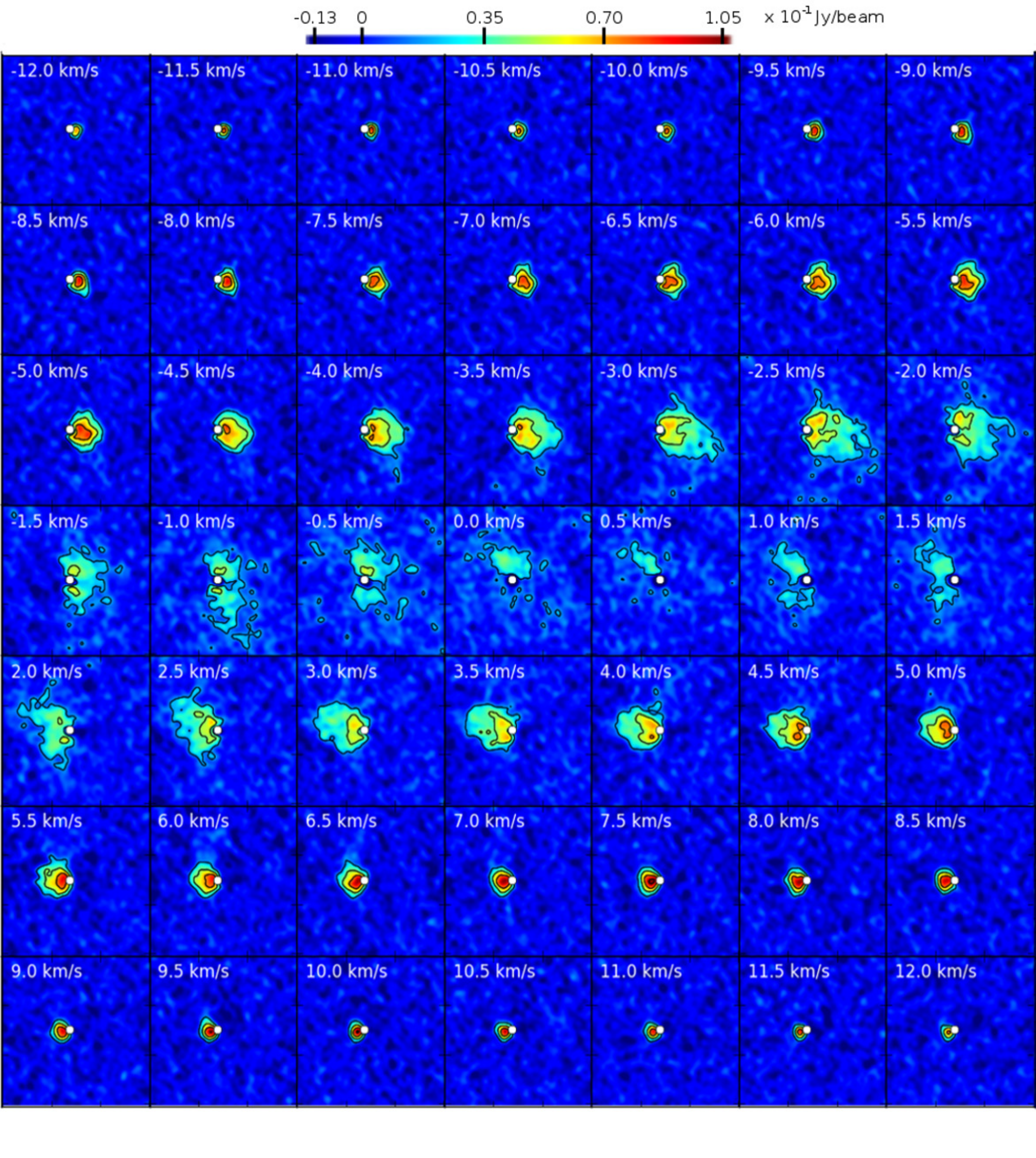}
\caption{Visualisation of the velocity channel maps of the $^{\rm 13}$CO $J=$3$-$2 emission of the circumstellar environment of L$_{\rm 2}$ Pup with the continuum subtracted. Each panel has a dimension of 1'' by 1''. The velocity channel is indicated in the top left corner of each panel. The black contour levels are drawn every 4 times the rms noise value outside the line. The white dot represents the location of the centre of mass. \label{chan_contsub2}}
\end{figure*}

\section{$^{\rm 13}$CO channel maps - radiative transfer model}

\begin{figure*}[h]
\centering
\includegraphics[width=\textwidth]{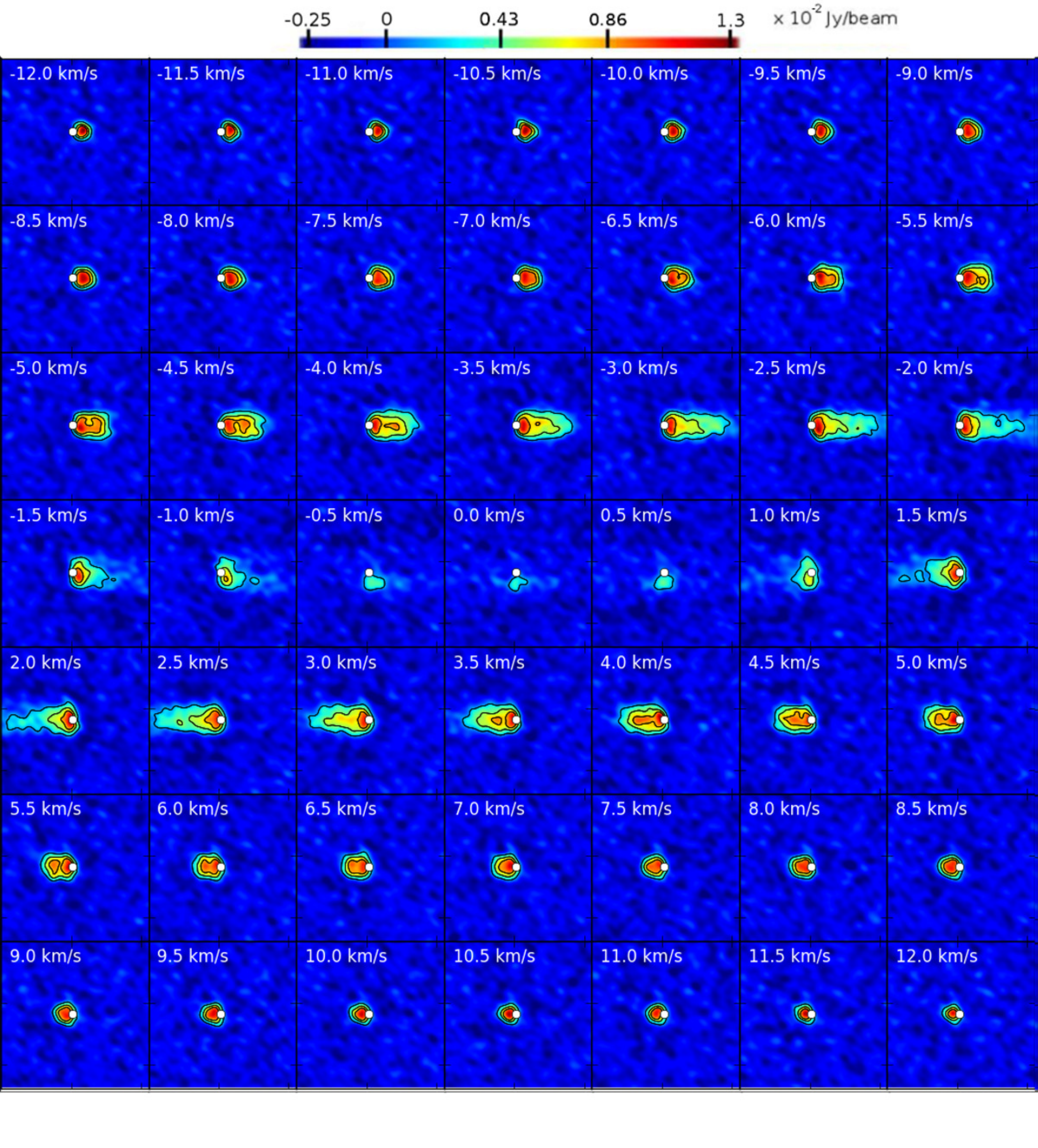}
\caption{Same as Fig. \ref{chan_contsub2}, but now showing the synthetic emission of the rotating disk model. \label{chan_model2}}
\end{figure*}

\end{appendix}
 
\end{document}